\documentclass[a4paper,11pt]{article}
\usepackage{amsmath, amssymb}
\usepackage{amsthm}
\usepackage{color, bm,graphicx}
\usepackage{cite}
\usepackage{slashed}

\theoremstyle{plain}
\newtheorem{definition}{Definition}
\newtheorem{theorem}{Theorem}
\newtheorem{lemma}{Lemma}

\newcommand{\email}[1]{\footnote{email:#1}}

\DeclareMathOperator{\ord}{ord}

\DeclareMathOperator{\tr}{tr}

\newcommand{\e}{\epsilon}
\newcommand{\G}{\Gamma^{(n)}}
\newcommand{\irr}{\text{irr}}
\newcommand{\fs}{\slashed}

\renewcommand{\O}{{\cal O}}

\renewcommand{\k}{{\underline k}}
\newcommand{\p}{{\underline p}}
\newcommand{\umu}{{\underline \mu}}

\newcommand{\be}{\begin{equation}}
\newcommand{\ee}{\end{equation}}

\begin{document}

\title{Self-Similar Structure of Loop Amplitudes \\and Renormalization}
\author{Kang-Sin Choi\email{kangsin@ewha.ac.kr}
 \\ \it \normalsize Scranton Honors Program, Ewha Womans University, Seoul 03760, Korea\\
\it \normalsize  Institute of Mathematical Sciences, Ewha Womans University, Seoul 03760, Korea }
\date{}
\maketitle

\begin{abstract}
We study the self-similar structure of loop amplitudes in quantum field theory and apply it to amplitude generation and renormalization. A renormalized amplitude can be regarded as an effective coupling that recursively appears within another loop. It is best described as a vertex function from the effective action. It is a scale-dependent, finite, parametrically small and observable quantity appearing in the $S$-matrix.
Replacing a tree-level coupling with a loop amplitude provides a systematic method of generating high-order loop amplitudes, guaranteeing no subamplitude divergence. This method also provides an alternative bottom-up proof to the traditional top-down recursive renormalization of general amplitudes.
\end{abstract}

\newpage
\tableofcontents

\section{Introduction}

Renormalization is about understanding the self-similarity of a system across varying scales. Its classic illustration is the block spin transformation, as discussed in Kadanoff's seminal work \cite{Kadanoff:1966wm}. In that, scaling up to include more spins allows for averaging them into a single spin, leading to a modified theory with the same formal structure but varying scale-dependent parameters. The scale is specified by temperature or site size. These modified parameters, known as ``renormalized'' physical parameters, are pivotal in renormalization, capturing its essence with varying scales. 

In this work, we employ the same idea to describe the processes involving elementary fields in quantum field theory (QFT). Physical parameters acquire scale dependence via renormalization and have a recursive structure.
A renormalized loop amplitude exhibits a recursive structure wherein it functions as a coupling constant within other amplitudes or recursively within itself. This recursive property enables the sum of such amplitudes to behave as an effective coupling analogous to the coupling observed in a free field theory.

Although the idea of self-similarity is well-established in statistical physics, it has yet to be fully utilized in particle physics. An exception is the famous running of the gauge couplings, which experiments have verified and led to the understanding of asymptotic freedom. The difficulty has been that in QFT, renormalization is mixed with the removal of infinities. Although loop corrections in QFT give rise to ultraviolet (UV) divergences, treating divergence may not be the essence of renormalization. 

The loop expansion is of a quantum nature because it accompanies powers of its characterizing Planck constant $\hbar$ \cite{Nambu:1968rr}. Moreover, the {\em recursive structure of the loop amplitudes from renormalizable couplings}, in quantum field theory, is the key to understanding self-similarity. The self-similarity of theory is reflected in perturbative corrections to physical parameters. Different scattering processes involving the same asymptotic states are indistinguishable and thus are superposed. As a result, amplitudes at various loop orders are added up or exchangeable. Also, inside an amplitude, another loop amplitude can be nested, and vice versa. 

Recent developments have rediscovered that a renormalized quantity expressed in terms {\em observables} is a meaningful physical one \cite{Choi:2023cqs, Choi:2023mma, Choi:2024hkd, Choi:2024cbs}. Crucially, in experimental settings, these renormalized parameters are what we observe, as they can be measured only by interactions with other particles and with our measurement apparatus, with finite external momenta \cite{Choi:2023cqs, Choi:2023mma}. This highlights a fundamental aspect: we should neither deal with the bare parameters in the defining Lagrangian nor the loop correction only, which are not separately observable in experiments. Suppose we use the physical parameters at a scale and the relative loop correction amplitude. In that case, we can define the same theory using the same number of finite parameters \cite{Choi:2024hkd, Choi:2024cbs}, and the theory is finite and perturbatively well-defined.

This recursive structure is well-encoded in the effective parameters, such as loop-corrected mass and $n$-point couplings, which are the coefficients of the effective action. These parameters span Green's function and are directly compared with the experimental data. 
The use of effective coupling eases the task of generating fully-renormalized higher-order amplitude systematically. 
The recursive generation of higher order amplitude is developed \cite{Kleinert:1982ki, Kastening:1999fy, Kleinert:1999uv, Ilderton:2005vg, Choi:2007nb, Kim:2008hda, Borinsky:2022lds, Abe:2009dr}. 
Polchinski's exact renormalization group \cite{Polchinski:1983gv} provides another method to reduce higher-order loops by integrating out high-loop operators by shrinking the loops. It showed that the vector space of the operator coefficients, in the $\phi^4$ theory, flows into three three-dimensional ones for kinetic, mass, and dimension four terms.

In this work, we generate the higher-loop amplitudes and renormalize them simultaneously. If we use the renormalized physical coupling to calculate the amplitude, we do not have subdiagram divergence; thus, the renormalization is automatic. It is crucial to deal with the renormalized propagator when doing this. 

Another important application is the full renormalization including the subamplitudes. It was completed by the program by Bogoliubov, Parasiuk, Hepp and Zimmermann (BPHZ) \cite{Bogoliubov:1957gp, Hepp:1966eg, Zimmermann:1967}, extended for the massless fields \cite{Lowenstein:1975rg, Lowenstein:1975ps} (see also\cite{BS, Collins:1984xc, Sibold, Blaschke:2013cba, Herzog:2017jgk, Itzykson:1980rh, Duncan:2012aja, Buchbinder:2021wzv, Talagrand:2022huy} for pedagogical introduction). This is a top-down approach that, from a given amplitude, tracks possible divergence of subamplitudes recursively. The Bogoliubov relation gives the relation between an amplitude and its subamplitude, and Zimmermann's forest formula provides the explicit solution. Our method provides an alternative bottom-up method and helps us understand how the forest formula works.

\paragraph{Organization} This paper is organized as follows. First, in Section 2, after taking examples in the famous $\phi^4$-theory using renormalization ``without counterterms,' \cite{Choi:2024cbs} we present the main idea. The regularization independence of the effective parameters enables us to use loop-corrected coupling in place of tree-level coupling. We take an example of generating two-loop correction of the propagator and the quartic coupling, using ``one-loop correction'' involving ``one-loop corrected effective coupling.'' This replacement is well-described in terms of coupling in the effective action. In Section 3, we establish the concepts appropriate to describe the self-similar nature of the amplitude and develop tools for investigating it. The self-similar nature is well described by the vertex functions, the coefficients of the effective action. The changeability of such couplings is defined in terms of the contraction. Irreducible amplitudes provide the minimal unit of such a process. In Section 4, we make use of this to generate general loop diagrams at arbitrary orders by iteration. Conversely, we can obtain the renormalization of a given amplitude of arbitrary order, which is done in Section 5. There, we prove the forest formula independently and provide a way to find a family of forests. We conclude with discussions in Section 6.

\paragraph{Notation} 
We denote a single $n$-point amplitude $A^{(n)}$ and the complete sum of $n$-point vertex function as $\Gamma^{(n)}$. The subscript of a physical quantity, for instance, $L$ in $\tilde \Sigma_L(p^2)$, denotes the loop-correction order, not the power of the coupling for the perturbation. To display its location $v$, the loop order $L$ and the weighted loop order $m$ and the momentum $\p$ dependence, we may denote it as $_mA_L^{(n)}(v,\p)$, but we usually keep it as brief as possible. We sometimes denote a graph with the same notation as the corresponding amplitude.   We make a convention to use the round bracket for accumulated loop orders, while we do not use the bracket for individual orders. That is, $\Gamma^{(n)}_{(L)}(\p)=\sum_{l=0}^L \Gamma^{(n)}_{l}(\p).$  From Section 3 on, for the renormalization of $A(\p)$, we use a bar as in $\overline A(\p)$.

\section{Two-loop amplitudes as effective one-loop ones} \label{sec:2loop}

We present our main idea in the simplest case: we show that two-loop amplitudes in $\phi^4$ theory are formally reduced to one-loop ones with ``effective coupling.''

\subsection{New Old Renormalization}

First, we briefly review renormalization ``without counterterms'' \cite{Choi:2024cbs}. In this paper, we are mainly interested in the $\phi^4$-theory, which appears as the Higgs sector in the Standard Model. The Lagrangian density is
\be \label{phi4Lag}
 {\cal L} = \frac12 \partial_\mu \phi \partial^\mu \phi - \frac12 m_B^2 \phi^2 - \frac{6\lambda_B}{24} \phi^4.
\ee
Here, the mass $m_B^2$ and the quartic coupling $\lambda_B$ are bare parameters defining the theory. They are defined in the absence of further interactions so that they are never directly observable. They receive quantum corrections so that the observed parameters are modified. 

Consider the quartic coupling $\lambda_B$. We have a one-loop corrected coupling
\be  \label{lambdacorr}
 \lambda_{(1)}(p_1,p_2,p_3,p_4) = \lambda_B Z^2 +6 \lambda^2 \left[ V(s) + V(t) + V(u) \right],
\ee
where 
\be
 i V(p^2) =\frac12 \int \frac{d^4k}{(2\pi)^4} \frac{i}{(p+k)^2-m^2+i\e} \frac{i}{k^2-m^2+i\e}. 
\ee
In what follows, we will omit the Feynman prescription $i \e$ for simplicity.
The factor $Z^2$ is reserved for the field strength renormalization that we see shortly. The symmetry factor $1/2$ is for the reflection symmetry of the internal loop.
We have three channels described by the Mandelstam variables 
\be \label{ManRel}
s=(p_1+p_2)^2, \quad t= (p_1+p_3)^2,\quad  u=(p_1+p_4)^2,\quad  s+t+u=4m^2.
\ee

\begin{figure}[t]
\begin{center}
\includegraphics[scale=0.6]{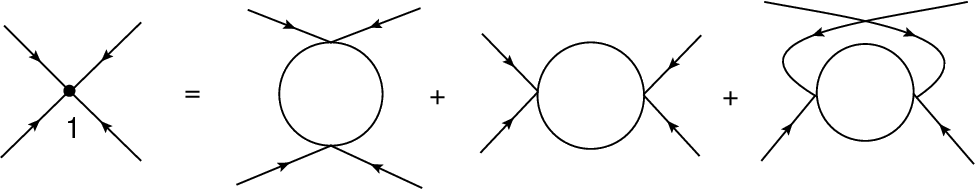}
\end{center}
\caption{One-loop correction for the quartic coupling in $\phi^4$ theory. The right-hand-side amplitudes after the renormalization are irreducible and their sum makes the effective coupling $\lambda_{1}(\p) = \Gamma_1^{(4),\irr}(\p)$, collectively depicted as the vertex on the left-hand side. The number around the resulting vertex denotes the loop order.\label{fig:lambda1}}
\end{figure}

We should be able to connect these results to observations.
We impose the renormalization condition at a reference scale $\mu_1,\mu_2,\mu_3,\mu_4$ satisfying $s_0\equiv(\mu_1+\mu_2)^2,t_0\equiv(\mu_1+\mu_3)^2,u_0\equiv(\mu_1+\mu_4)^2,s_0+t_0+u_0=4m^2$ such that the scattering matrix element ${\cal M}$ for four scalars becomes the coupling $\lambda$ 
\be \label{LambdaRenCond} \textstyle 
 -6 \lambda_{(1)} \left(\mu_1,\mu_2,\mu_3,\mu_4\right)= {\cal M}\left(\text{4 scalars}\right) \equiv -6 \lambda + \O(\hbar^2 \lambda^3).
\ee
This choice of using the physical parameter is called the on-shell (OS) scheme \cite{Sirlin:1980nh}, which we take in this paper.
This gives us the relation between $\lambda$ and $\lambda_B Z^2$
\be \label{lambda1match} \textstyle
6\lambda = 6 \lambda_{(1)}\left(\mu_1,\mu_2,\mu_3,\mu_4\right) =  6\lambda_B Z^2 + 36\lambda^2 \left[ V(s_0)+V(t_0)+V(u_0)\right] + {\cal O}(\hbar^2 \lambda^3).   
\ee
Note that this is nothing more than the conventional separation of the bare coupling into the physical and the counterterm parameters.
These two equations enable us to remove the bare parameter $\lambda_B Z^2$ to give 
\be \label{lambda1wo}
\begin{split}
6\lambda_{(1)}(p_1,p_2,p_3,p_4)&= 6\lambda +36\lambda^2 \Big( \left[  V(s) - V(s_0)] + [V(t) -V(t_0)] +[ V(u)  - V(u_0) \right] \Big) \\
&=6 \lambda +\frac{36 \lambda^2}{32\pi^2} \int_0^1 dx \log \frac{m^2-x(1-x)s}{m^2-x(1-x)s_0} + (s \leftrightarrow t)+(s \leftrightarrow u) .
\end{split}
\ee
This rewriting of loop-corrected coupling is the {\em renormalization} in the conventional sense. The value $\lambda_{(1)}(p_1,p_2,p_3,p_4)$ is unchanged, which is now expressed as the {\em relative} coupling in the unit of the reference coupling $\lambda$ defined in (\ref{LambdaRenCond}). In this sense, we will call this relative-value rewriting (RVR). 

Although $V(s)$ alone is divergent, but the combination $V(s)-V(s_0)$ is not, and hence $6\lambda_{(1)}(p_1,p_2,p_3,p_4)$ is finite as well. The high-momentum modes in (\ref{lambdacorr}) are insensitive to the external momenta and are canceled in (\ref{lambda1wo}). The result depends on the two scales, $s$ and $s_0$.  It is regularization-independent because it is finite \cite{Zimmermann:1969jj}. We may verify that any choice of regularization gives the same result \cite{Choi:2023cqs}. It is small and admits perturbative expansion with the small $\lambda$. 
It has well-defined loop corrections
\be
 \lambda(\p) = \lambda + \sum_{l=1}^\infty \lambda_l (\p),
\ee
where $\lambda_l(p^2)$ are the $l$-loop corrections
\begin{align}
 6\lambda_0(\p) &= 6\lambda \label{lambda0} \\
 6\lambda_1(\p) &= \hbar (6\lambda)^2 \sum_{\{(p^2,p_0^2)\}} \big( V(p^2)-V(p_0^2) \big), \label{lambda1} 
\end{align}
where the sum takes place over $(p^2,p_0^2)=(s,s_0),(t,t_0),(u,u_0).$
In fact, for this reason, we used the coupling $\lambda$ instead of the bare one in calculating the loop correction (\ref{lambdacorr}). With higher order results, thus, the matching condition like (\ref{lambda1match}) (re)defines the coupling $\lambda$ more accurately.

Next, the mass is similarly corrected by the self-energy $\tilde \Sigma(p^2)$. The quadratic interaction becomes
\be \label{corrmass}
 p^2 - m_B^2 - \tilde \Sigma (p^2)
\ee
We need a reference mass for expanding the theory.
One natural choice is the pole mass $m$, defined as\footnote{Taking the pole mass means adopting the OS scheme. We do it only for the masses.}
\be \label{scpolemass}
 \left[ p^2 - m_B^2 - \tilde \Sigma (p^2)\right]_{p^2=m^2} = 0,
\ee
Using this, we can eliminate the bare mass and express the corrected quadratic interaction (\ref{corrmass}) as $p^2 - m^2 - \tilde \Sigma (p^2) + \tilde \Sigma (m^2)$.
We define the propagator as its inverse
\be \label{CorrProp}
 D'(p^2)=  \frac{i}{p^2 - m^2 - \tilde \Sigma (p^2) + \tilde \Sigma (m^2)},
\ee
whose pole is located indeed at $p^2=m^2$, justifying the name of the reference mass.
Due to the correction, the residue of the propagator at this point is different from that of the free field
\be  \label{porpres}
 p^2 \to m^2:  D'(p^2) \to  \frac{iZ}{p^2-m^2} + {\cal O}((p^2-m^2)^2), \quad Z^{-1} = 1- \frac{ d \tilde \Sigma}{d p^2}(m^2),
\ee
which comes from the next leading order expansion from $\tilde \Sigma(p^2)$. 

We demand a self-similarity such that the corrected propagator (\ref{porpres}) effectively behaves as a free field around the expansion point. So, we re-normalize the field, called field-strength re-normalization (FSR) 
\be \label{FSren}
 \phi \equiv \sqrt{Z} \phi_r.
\ee 
The resulting propagator for this $\phi_r$ is
\be \label{FeynProp}
 D(p^2) = \frac{i}{p^2-m^2(p^2)},
\ee
where we are naturally led to define the momentum-dependent mass \cite{Coleman:2018mew}, to ${\cal O}((p^2-m^2)^2)$
\be \label{SlidingMass}
\begin{split}
m^2(p^2)& = m^2 + \tilde \Sigma(p^2) - \tilde \Sigma(m^2) - (p^2 -m^2) \frac{ d \tilde \Sigma}{d p^2} (m^2) \\
 & \equiv m^2 + \big(1 - t^2_m \big) \tilde  \Sigma(p^2).
\end{split}
\ee
As a result, this has the form of the Taylor expansion. We denote its operator by $t^2_m$, in which the superscript indicates the expansion up to the second order and the subscript does the renormalization point $p^2=m^2$. The linear term in $p$ is absent due to the symmetry $p \leftrightarrow -p$. 
As in the case of the quartic coupling (\ref{lambda1wo}), we may check that the combination $\big(1 - t^2_m \big) \tilde \Sigma(p^2)$ is finite \cite{Zimmermann:1969jj} and parametrically small in $\lambda$ \cite{Choi:2024cbs,Choi:2024hkd}. At the leading order, 
\be \label{FeynProp0}
 D_0(p^2) = \frac{i}{p^2-m^2}.
\ee 
It turns out to be tree-level or zero-loop order, so we put the subscript 0. 

The success of QFT is the renormalization in the Kadanoffian sense: the prediction of the {\em scale dependence of the physical parameter.} The energy-dependent quantity $m^2 (p^2)$ plays the role of the mass as in the free field. We call this self-similar renormalization (SSR). We can measure it from the reference mass $m^2$ without using the bare mass $m_B^2$ \cite{Choi:2023cqs} by scattering experiments. As before, the loop corrected mass (\ref{SlidingMass}) is the same, just re-expressed in terms of finite quantities; the additional terms appeared because this is now the mass of the re-normalized quantity $\phi_r$ in (\ref{FSren}). We specify the scale by the {\em external momentum $p^2$} that the experiment can measure. The parameters $\lambda_{(1)}(\mu_1,\mu_2,\mu_3,\mu_4), m^2(p^2)|_{p^2=m^2}$ are not the predictions, although they are defined by renormalization, but the inputs or the ``boundary conditions.'' In reality, what we have observed from the running gauge coupling is its dependence on the external momentum. If one insists, we may generalize the concept of the mass as that in (\ref{SlidingMass}), however, it is not separately observable. It is clear that only the whole combination in the propagator is physical, and it is not meaningful to mention a part of (\ref{FeynProp}).

\begin{figure}[t]
\begin{center}
\includegraphics[scale=0.7]{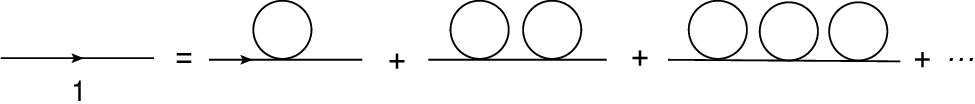}
\end{center}
\caption{The weighted Feynman diagram for the one-loop corrected propagator in $\phi^4$ theory. Although the two-point vertex function receives one-loop correction $\Gamma^{(2)}(p^2) = p^2 - m^2 - (1-t^2_m) \overline \Gamma^{2}(p^2) + {\cal O}(\hbar^2)$, the propagator is expanded as the infinite sum $D(p^2) = 1/\Gamma^{(2)}$. In the $\phi^4$ theory, all the corrections vanish. \label{fig:M1}}
\end{figure}

The one-loop corrected propagator diagram is drawn in Fig. \ref{fig:M1}. The one-loop correction is
\be \label{Sigma1}
\begin{split}
 - i \tilde \Sigma_{1}(p^2) &=\frac12 \int \frac{d^4 k}{(2\pi)^4} (-6i \lambda) D_0(k^2) \\
&=  -\int \frac{d^4 k}{(2\pi)^4} \frac{3i\lambda}{ k^2 - m^2} .
\end{split}
\ee
It is independent of the external momentum. Thus, from (\ref{SlidingMass}), it is renormalized to zero
\be
 m_1^2(p^2) = (1-t^2_m) \tilde \Sigma_{1}(p^2) =0.
\ee
It is a well-known specialty in the $\phi^4$ theory and spontaneously broken non-Abelian gauge theories. Otherwise, we may have this loop inside the general loop, which cannot be removed by hand \cite{Choi:2024cbs}. The nontrivial contributions arise from the two-loop order that we calculate below.

\subsection{Two-loop propagator from effective couplings}

\begin{figure}[t]
\begin{center}
\includegraphics[scale=0.7]{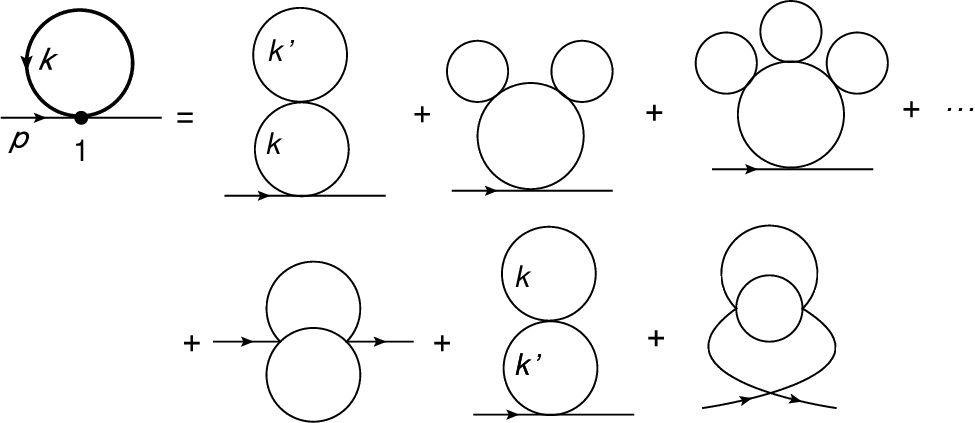}
\end{center}
\caption{Two-loop propagators are obtained from a one-loop one by replacing the quartic coupling with the physical one-loop corrected. The effect of the one-loop corrected propagator $m^2_1$ is on the first line, and that of the one-loop corrected vertex $\lambda_1$ is on the second line. The latter receives contributions from $s,t,u$-channels. Note that there can be identical diagrams from different origins. \label{figM2}}
\end{figure}

A loop amplitude that depends on the external momenta can play the role of the {\em effective coupling.} 
We can formally write down a ``one-loop'' amplitude using this. In place of one of the tree-level couplings $\lambda$, we may use the one-loop correction $\lambda_{1}(\p,k)$ in (\ref{lambda1wo}),
\be \label{FormalOneloop}
  - i \tilde \Sigma_1(p^2) \Big|^1_{\lambda \to \lambda_1(\p,k)} = \frac12 \int \frac{d^4 k}{(2\pi)^4} D_0(k^2)  [-6i \lambda_{1}(p,-p,-k,k)] .
\ee
The superscript 1 indicates that the replacement takes place at one place.
Letting $p$ be the external momentum, the momenta $(p,-p,-k,k)$ flow in the vertex, summing to zero. This is the case not only at the vertex for the tree-level coupling $\lambda$ but for the loop amplitude $\lambda_{1}(\p,k)$.
From  (\ref{lambda1wo}), the coupling $\lambda_{1}(\p,k)$ has three channels $s=(p+k)^2,t=0,u=(p-k)^2$. Thus the equation (\ref{FormalOneloop}) means that
\be \label{FormalOneloopExp} 
 \tilde \Sigma_1 (p^2) \Big|^1_{\lambda \to \lambda_1(\p,k)}=\frac{(6\lambda)^2}2 \int \frac{d^4 k}{(2\pi)^4}D_0(k^2) [ V(s)-  V(s_0)  + V(t) -  V(t_0) +  V(u) -  V(u_0)] .
\ee
This becomes the sum of amplitudes at {\em two-loop} order with renormalized subdiagrams. The corresponding diagrams before renormalization are the last three diagrams in Fig. \ref{figM2}. 
 This time $\lambda_1(p,-p,-k,k)$ depends on $p^2$ through the dependence on $V(s)$ and $V(u)$.

The $s$- and $u$-channel amplitudes in (\ref{FormalOneloopExp}) are the famous ``sunset'' diagrams in which the one-loops are {\em renormalized.} In each, a potential ``subdiagram'' divergence in the first term is canceled by the second term, regarded as the counterterm contribution. The $t$-channel contribution is zero because $V(t)=V(t_0)$ is independent of the external amplitude. 

The sunset diagram is famous because of the overlapping internal lines shared by different loops.
Our construction has two interesting points. The corresponding diagram has ``enhanced'' symmetry, that is, the shuffling among three internal loops.
This is unexpected because the one-loop $\lambda_1(\p,k)$, which locally replaces the tree-level coupling, does not see the global structure of the resulting diagram. The sunset diagram has $3!$ permutation symmetry among the loops. So, we modify the symmetry factor by multiplying $1/3!$. If all three internal propagators are distinguishable, which can happen, for instance, if we had more than two fields with ``generic'' interactions, we do not need to introduce the symmetric factor by hand. Conversely, if we are given a sunset diagram, we know that those $3!$ amplitudes are identical and there is only one kind of divergence. 

It is the second interesting point that the multiplicity along the time ordering direction is automatically taken into account in the full amplitude. The $s$-channel amplitude alone does not have the flipping symmetry the original one-loop diagram has. However, the $u$-channel diagram is the desired flipped diagram with the same amplitude so that the sum correctly cancels the symmetry factor $1/2$ of the one-loop. We have
\be \label{FormalOneloopExp2}
\begin{split}
 \tilde \Sigma_{2}(p^2) & = \frac{1}{3!} \tilde \Sigma_1(p^2) \Big|^1_{\lambda \to \lambda_1(\p,k)} \\
  &=\frac{(6\lambda)^2}{2 \cdot 3!} \int \frac{d^4 k}{(2\pi)^4} [ V((p+k)^2)-  V(s_0) +  V((p-k)^2) -  V(u_0)] D_0(k^2) \\
 & =\frac{(6\lambda)^2}{3!} \int \frac{d^4 k}{(2\pi)^4} [ V((p+k)^2)- \frac12 V(s_0)- \frac12 V(u_0) ] D_0(k^2).
\end{split}
\ee
The constant term will be canceled after the full renormalization so that replacing $V(u_0)$ with $V(s_0)$ gives the same answer. 

Naturally, the loop structure induces nontrivial momentum dependence in $\lambda_{(1)}(\p,k)$, so we have more than one correction amplitude. However, the total amplitude, being Lorentz scalar in $p$, only depends on the size of the external momentum $p^2$ after performing the integration over $k$. We know this using the symmetry $k \leftrightarrow -k$, it is an even function in $p$, $ \tilde \Sigma_{(2)}(-p)= \tilde \Sigma_{(2)}(p)$ and depends on the even powers of $p$. Indeed, we have
\be
 \tilde \Sigma_{2}(p^2) = \frac{3i\lambda^2}{32\pi^2} \int_0^1 dx \int \frac{d^4k}{(2 \pi)^4} \frac{1}{k^2-m^2} \log \frac{F_{k^2}(p^2)}{(m^2-x(1-x)s_0)(m^2-x(1-x)u_0)},
\ee
with a $p^2$-dependent function
$$
 F_{k^2}(p^2)= \left(m^2-x(1-x)(p^2-k^2)\right)^2-4x^2(1-x)^2m^2k^2.
$$

We may consider similar contributions from the propagator correction
\be
  \frac12 \int \frac{d^4 k}{(2\pi)^4} [-6i \lambda_0(p,-p,-k,k)]\frac{i}{k^2-m_{(1)}^2(k^2)},
\ee
which gives rise to further loop diagrams in the second line in Fig. \ref{figM2}. However, they do not depend on the external momenta and thus are all zero. 

The overall amplitude is renormalized as in (\ref{SlidingMass}), so that 
\be \label{SlidingMass2} \begin{split}
 m^2_{(2)}(p^2) &= m^2 +(1-t^2_m)  \tilde \Sigma_{(2)}(p^2)  \\
& = \sum_{l=0}^2 m_{l}^2 (p^2),
\end{split}
\ee
where
\begin{align}
 m_{0}^2(p^2) &= m^2, \\
 m_{1}^2(p^2) &= 0, \\
 m_{2}^2(p^2) &= \frac{ 3 i \lambda^2}{32\pi^2}\int_0^1 dx \int \frac{d^4k}{(2\pi)^4} \frac{1}{k^2-m^2}  \left[ \log  \frac{F_{k^2}(p^2)}{F_{k^2}(m^2)} - (p^2-m^2)\frac{dF_{k^2}/dp^2 (m^2)}{F_{k^2}(m^2)}  \right].
\end{align}
The constant terms in the square bracket in (\ref{FormalOneloopExp2}) are canceled, as it should be. The diagrams for these terms have the same structure as that of the previous one-loop, in the first line in Fig. \ref{figM2}, which are renormalized away. Note that in our scheme the mass $p^2=m^2$ is physical and unchanged $m_{(2)}^2(m^2) = m^2$. However, the effective mass has a nontrivial profile in $p^2$ away from it.
Accordingly, we can define a similar propagator $D_{(2)}(p^2)$ up to two loops.
Note that, although $m_1^2(p^2)$ vanishes due to renormalization, $m_2^2(p^2)$ made out of $m_1^2(p^2)$ by the replacement does not, because it acquires the dependence on the external momentum.

\subsection{Two-loop quartic coupling from effective couplings} \label{sec:2loopquartic}

\begin{figure}[t]
\begin{center}
\includegraphics[scale=0.8]{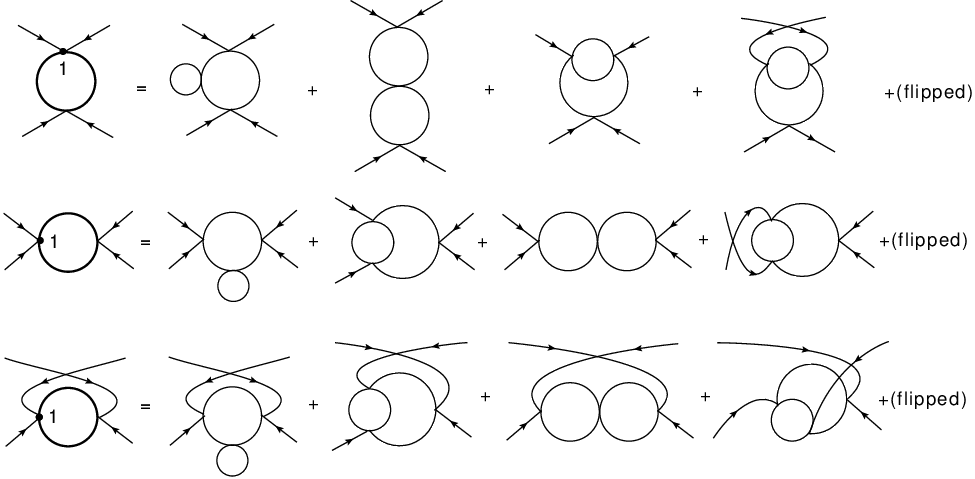}
\end{center}
\caption{Two-loop corrections for the quartic coupling in $\phi^4$ theory using weighted one-loop Feynman diagrams. We obtain them by replacing one of the vertices in a one-loop diagram ($s,t,u$ respectively, rows) with the one-loop renormalized $\lambda_{1}(\p)$ $s,t,u$-vertices (columns). Replacing the other vertices gives similar diagrams, which are line-reflected diagrams of these. All the subdiagram parts corresponding to $\lambda_{1}(\p,k)$ are renormalized and finite. \label{fig:lambda2}}
\end{figure}

In the same way, we calculate the two-loop corrections to the quartic coupling using the one-loop correction $\lambda_1(\p)$ with replacement,
\be \label{FormalQuartic} 
\begin{split}
6\lambda_1(\p) \Big|^1_{\lambda \to \lambda_1(\p,k)}  &= \frac{1}{2 \cdot 2} \int \frac{d^4k}{(2\pi)^4} [-6 i \lambda_{1}(\p,k)] 
 D_0((p+k)^2) D_0(k^2)(-6 i\lambda) \\
  &\quad + \frac{1}{2 \cdot 2} \int \frac{d^4k}{(2\pi)^4} (-6 i\lambda) D_0((p+k)^2) D_0(k^2) [-6 i \lambda_{1}(\p,k)].
\end{split}
\ee
As before, the superscript 1 indicates that the replacement is done one by one.
Here and hereafter, we denote the set of external momenta as $\p$ without confusion. The symmetric factor $1/2$ is that of one-loop. The other factor $1/2$ came from the exchange symmetry between the two vertices, giving the same amplitude; if we have more different scalar fields with generic interactions, $\lambda_{1}$'s would have been distinguishable in both lines in (\ref{FormalQuartic}), and so have been the resulting two-loop diagrams. 

Consider the first line in (\ref{FormalQuartic}) first. 
Each vertex lies inside the loop with the internal momentum $k$, to which four momenta $(\p,k) \equiv \{-k,-p_i,-p_j,k+p_i+p_j\}$ flow in, preserving the total momentum. Here, $p_i$ and $p_j$ $(i\ne j)$ originate from the channel of the original one-loop diagram as in (\ref{ManRel}). Each vertex also has three independent channels from $\p$. Therefore, we have three possible channels in the subdiagram as functions of $s'=(p_i+k)^2, t'=(p_j+k)^2, u'=(p_i+p_j)^2$, respectively:
\begin{align} 
\begin{split} 
 6\lambda_{2}(\p) &= 6\lambda_1(\p) \Big|^1_{\lambda \to \lambda_1(\p,k)} \\
 & \equiv (-6i\lambda)^3 \big( W_s(s) + W_t(t) + W_u(u) \big) , 
 \end{split} \\
\begin{split} 
 W_s(s) &= \frac12 \int \frac{d^4k}{(2\pi)^4}D_0((p_3+p_4+k)^2)D_0(k^2)  \\
 & \qquad \times \Big(  \left[  V(s)-  V(s_0)\right] +\left[ V((p_4+k)^2) -  V(t_0)\right]   + \left[ V((p_3+k)^2) -  V(u_0) \right] \Big) ,
\end{split} \label{Wschannel}
\\
\begin{split}
  W_t(t) &= \frac12 \int \frac{d^4k}{(2\pi)^4} D_0((p_1+p_3+k)^2) D_0(k^2) \\
 & \qquad \times \Big(  \left[  V((p_1+k)^2)-  V(s_0) \right] + \left[ V(t) - V(t_0) \right]  + \left[ V((p_3+k)^2) -  V(u_0) \right] \Big)  ,
\end{split}
\\ \begin{split}
 W_u(u) &= \frac12 \int \frac{d^4k}{(2\pi)^4}  D_0((p_1+p_4+k)^2)D_0(k^2) \\
 & \qquad \times \Big(  \left[  V((p_1+k)^2)-V(s_0) \right] + \left[ V((p_4+k)^2) -V(t_0) \right]  + \left[ V(u) - V(u_0)  \right]\Big).
\end{split}
\end{align}
We have two-loop corrections as a result. We may also replace the propagators $6\lambda_1(\p) \big|^1_{D_0(\p,k) \to D_1(\p,k)}$.
The corresponding diagrams are drawn in Fig. \ref{figM2}. 

They are arranged in terms of the Mandelstam variables, which are the only scalar quantities made of the conserved momentum. It is highly nontrivial from the asymmetric dependence on the external momenta to see that each integral is a function of Mandelstam variables.
For instance, in (\ref{Wschannel}), it is known that the finite parts of the second and the third bracketed terms are functions of $s$ only \cite{Peskin:1995ev, Kleinert:2001ax}. Thus, they are the same. This desirably cancels the symmetric factor that existed in the one-loop.\footnote{All of these corrections are expressed by the single function $W(p^2)=W_s(p^2)=W_t(p^2)=W_u(p^2)$, if we take the renormalization point symmetrically $s_0=t_0=u_0=4m^2/3.$
Indeed, we can verify that every unrenormalized diagram matches another in each row after changing the order of the external momenta.} 

We have another set of contributions, whose diagrams are obtained by flipping the ones in Fig. \ref{figM2}. We may regard the latter as replacing the other $\lambda$ by $\lambda_{1}(\p,k)$, which is the second line in (\ref{FormalQuartic}). They are equal. 

Therefore, we have obtained the quartic couplings at two-loop
\be
 \lambda_{(2)}(\p) = \lambda_B Z^2 + \sum_{p^2=s,t,u} \big[ 6 \lambda^2 V(p^2) + 36 \lambda^3  W (p^2) \big].
\ee
Again, there is no subdiagram divergence because all the sub-loops contained in the modified coupling $\lambda_{1}(\p,k)$ are renormalized.
The renormalization condition for the total amplitude is the same as before in (\ref{LambdaRenCond}),
\be \label{lambda2match}
\begin{split}
 \lambda = \lambda_{(2)}(\umu) 
 = \lambda_B Z^2 + \sum_{p_0^2= s_0,t_0,  u_0}  \big[  6 \lambda^2 V(p_0^2) + 36 \lambda^3 W(p_0^2) \big]+ \O(\hbar^3\lambda^4).
\end{split}
\ee
In other regularization scheme other than OS, we need to {\em match again} the bare coupling $\lambda$, valid at ${\cal O}(\hbar^2 \lambda^3)$, instead of putting the relation (\ref{lambda1match}), which is not accurate enough to this order. It is because in calculating $V(p^2)$ and $W(p^2)$, which also depend on $\lambda$, we have used more precise coupling $\lambda$ satisfying the equation (\ref{lambda2match}), which is different from that in (\ref{lambda1match}) (otherwise the order-2 correction should be zero).

The renormalization of the overall amplitude needs some care. To see this, let us separate the amplitude (\ref{Wschannel}) as
\begin{align}
 W_s(s) & \equiv W_8(s)+W_\circledcirc(s),\\
  \begin{split}
 W_8(s) & \equiv  \frac12 \left[  V(s)-  V(s_0)\right]  \int \frac{d^4k}{(2\pi)^4} D_0((p_3+p_4+k)^2)D_0(k^2) 
 \end{split}   \label{w8}
 \\
 \begin{split}
 &= \left[  V(s)-  V(s_0)\right]  V(s), \\
 W_\circledcirc (s) & \equiv \frac12 \int \frac{d^4k}{(2\pi)^4} D_0((p_3+p_4+k)^2)D_0(k^2)  \\
 & \qquad \times  \Big( \left[ V((p_4+k)^2) -  V(t_0)\right]  + \left[ V((p_3+k)^2) -  V(u_0) \right]\Big) .
\end{split}
\end{align}

We have the unique renormalization, so we use the same reference scale $p^2=s_0, t_0,u_0$ as before. However, $W_8(s_0) =0$ and its subtraction from $W_8(s)$ does not make the amplitude finite.
The reason is that the newly inserted one-loop $V(s)$ does not share any internal line with the original loop. The two subloops are {\em disjoint}\footnote{They are not strictly disjoint but share the vertex, however the renormalization comes from the loop structure.} and do not affect the divergence structure of each other. That is, no other structure of divergence appears than the one-loop one before insertion. Therefore, the limiting procedure (\ref{lambda2match}) renormalizes the ``original'' part only:
\be
 \begin{split}
\overline W_8(s) &=  \frac12 \left[  V(s)-  V(s_0)\right]   \\ 
&\quad  \times\frac12 \int \frac{d^4k}{(2\pi)^4}  \Big( D_0((p_3+p_4+k)^2)D_0(k^2) -D_0((p_3+p_4+k)^2)D_0(k^2)_{s=s_0} \Big) \\
 &= \left[  V(s)-  V(s_0)\right]  \left[  V(s)-  V(s_0)\right]. 
 \end{split}
\ee
This is desirable because the amplitude has the exchange symmetry between the two subloops. The renormalization {\em already done at one-loop level} should take care of it. There is no divergence belonging to only the total diagram, which is a two-loop, and there is the ``counterterm,'' say $\delta_{\lambda2s}\phi^4/24$, satisfying a redundant relation
\be \label{disjct}
 \frac{1}{24} \delta_{\lambda2s} \phi^4 = \frac{6\lambda}{24} (6\lambda V(s_0))^2 \phi^4.
\ee
If this relation is not satisfied, it will give rise to another two-loop with the same shape as (\ref{w8}) with all the vertex couplings $\delta_{\lambda2s}$, which must be canceled by another counter-counterterm; this would repeat infinitely and the theory cannot be renormalized by the finite number of counterterms \cite{Choi:2024cbs}. We check that there are similar amplitudes in the $t$ and $u$ channels.

For $W_\circledcirc(s)$, the newly inserted loop $V((p_4+k)^2)$, etc., is nested in the total two-loop $W_\circledcirc(s)$. Those $V((p_4+k)^2)$, etc., depend on the momentum $k$ of the main amplitude $W_\circledcirc$ from which the divergence appears. In this case, we renormalize the overall amplitude as before $\overline W_\circledcirc(s) = W_\circledcirc(s)-W_\circledcirc(s_0)$.

Thus, we add the physical coupling at two-loop order on top of (\ref{lambda0}) and (\ref{lambda1})
\be \label{Gamma42} 
  6\lambda_{2} (\p)= \hbar^2 (6 \lambda)^3  \sum_{\{(p^2,p_0^2)\}} \big( W_\circledcirc(p^2)- W_\circledcirc (p_0^2) +\big [V(p^2)-V(p_0^2) \big]^2\big), 
\ee
with the same sum range as in (\ref{lambda1}).
Again, the RVR renormalization is rewriting of the same coupling: the bare parameter $\lambda_B$ has disappeared and the additional loop correction at $\O(\hbar^2 \lambda^3)$ is given by the difference of the quantity $W(\p)$ at different scales, removing the dependence on the cutoff. 

Note that we may change the order of the insertion and the renormalization if we obey the rule for the disjoint subamplitudes. The total amplitude $\lambda_{(2)}$ is the same, while the corrections in terms of the renormalized quantities are always given as the difference as in (\ref{Gamma42}). In the rest of the paper, we replace couplings with the renormalized subamplitudes inside the renormalized amplitude.

\section{Self-similar structure}

In the previous section, we have generated two-loop couplings $\lambda_{2}(\p)$ and $m^2_{2}(\p)$ as the ``one-loop'' amplitude by replacing the coupling to one-loop amplitude $\lambda_{1}(\p)$. Such replacement is best defined in terms of the effective couplings and the coefficients of the effective action. 

\subsection{Effective couplings}

The renormalized couplings and propagators are best understood as vertex functions in the momentum space. Following the standard functional method \cite{Jackiw:1974cv}, we have the generating functional through the path integral
\be \label{genfnal}
 Z[J] = e^{-i E[J]} = \frac{1}{Z[0]} \int {\cal D}\phi \exp \left[ \frac{i}{\hbar} \int d^4x \left( {\cal L}[\phi]+J \phi \right) \right].
\ee
Here, ${\cal L}[\phi]$ is the Lagrangian defining the theory. 
Evaluating the path integral over $\phi$  gives the free energy $E[J]$ as an effective action containing the quantum corrections.
Defining the classical field $\overline \phi$ and its quantum fluctuation $\hat \phi$
\be
 \overline \phi(x) \equiv - \frac{\delta E[J]}{\delta J(x)} , \quad\phi(x) \equiv \overline \phi(x) + \sqrt{\hbar} \hat \phi(x),
\ee
the Legendre transform replaces the dependent variable $J$ with $\overline \phi$, yielding the {\em usual} effective action
\be
 \Gamma[\overline \phi] = - E[J] - \int d^4x J(x) \overline \phi(x), \quad \quad \frac{\delta \Gamma[\overline \phi]}{\delta \overline \phi} = -J.
\ee
The effective action is expanded in the classical fields $\overline \phi(x)$, which defines $n$-point vertex functions
\be \label{effaction}
 \Gamma^{(n)}(x_1,\dots, x_n) = \frac{\delta^n \Gamma[\overline \phi] } {\delta \overline \phi(x_1) \dots \delta \overline \phi(x_n)} \bigg|_{\overline \phi=0}. 
\ee
In this paper, we consider a renormalizable Lagrangian, in which operators have dimensions less than or equal to four. Nevertheless, we may have high-order interactions $n>4$ in the effective action.

It is well-known that the $n$-point effective action is the same as the 1-particle irreducible $n$-point correlation function
\be
 \Gamma^{(n)}(x_1,\dots,x_n) = -i \langle \phi (x_1) \dots \phi (x_n) \rangle_{\text{1PI}}
\ee
that completely expands the theory \cite{Peskin:1995ev}. Also, this is the minimal unit that appears in the $S$-matrix. (In fact, we need further field-strength renormalization for quantities with $\delta>0$, see below.) Therefore, we can use the effective action that defines the theory. 

We use the coupling in the momentum space
\be \label{coupling}
(2\pi)^4 \delta^{(4)}(p_1+ \dots + p_n) \Gamma^{(n)}(\p) = \int \frac{d^4 p_1}{(2\pi)^4} \dots \frac{d^4 p_n}{(2\pi)^4} \Gamma^{(n)}(x_1,\dots, x_n) e^{-i( p_1\cdot x_1 + \dots + p_n\cdot x_n)},
\ee
where $(\p)$ is a shorthand for $(p_1,p_2, \dots, p_n)$. 

The {\em momentum-dependent coupling} is expanded in loop order
\be \label{Gammaexp}
 \hbar^{-1} \Gamma^{(n)} (\p) = \hbar^{-1} g^{(n)}_B + \hbar^{-1}  \sum_{L=1}^\infty \Gamma^{(n)}_L (\p),\quad \Gamma^{(n)}_L (\p) \propto \hbar^L,
\ee
where $\Gamma^{(n)}_l (\p)$ is the $l$-loop correction and the zeroth order coupling is the bare parameter in the Lagrangian \cite{Jackiw:1974cv}. An obvious but important property shared by all order couplings is momentum conservation
\be \label{momconserv}
 p_1+ \dots + p_n = 0, \quad \text{ for all $l$}.
\ee
For the zeroth order $L=0$, it is conservation at the vertex. For higher order, it means the sum of the inflowing momenta into the {\em loop} is zero.
We justify the $\hbar$-dependence in (\ref{Gammaexp}) by the following theorem.
\begin{theorem}[(Nambu, Coleman, Weinberg \cite{Nambu:1968rr}) loop expansion is in $\hbar$] \label{thmloop}
The loop expansion parameter is $\hbar$. That is, for the number of loop $L$, we have
\be \label{loophbar}
 \hbar^{-1} \G_L(\p) \propto \hbar^{ I - n} = \hbar^{L-1}. 
\ee
\end{theorem}
From the overall factor $\hbar^{-1}$ of the action (\ref{genfnal}), we see that each coupling $\hbar^{-1} \G$ carries $1/\hbar$. 
The propagator in the amplitude $(\hbar^{-1} \Gamma^{(2)})^{-1}$ carries $\hbar$, whose number $I$ is that of lines in the diagram. From the amplitude, the order $n$ coupling $\hbar^{-1} \G(\p)$, having $n$ vertices and $I$ propagators, is proportional to $\hbar^{I-n}$. Also, for each loop, we have the momentum conservation (\ref{momconserv}) at each vertex, so the number of loops $L$ is counted by the number of internal lines \footnote{This is Euler characteristic in the diagram side, which is regarded as a polygon projected on a surface thus the relation (\ref{Euler}) is the Euler number of polyhedra with one face removed.}
\be \label{Euler}
 L = I - n + 1,
\ee
the number of loops remains unchanged. Here, $+1$ is from the overall momentum conservation, overcounted at each vertex. We obtain the relation (\ref{loophbar}) yielding the desired one (\ref{Gammaexp}).

An important exception is that the bare coupling is not necessarily the zeroth order in (\ref{Gammaexp}). Indeed, it is not. The zeroth order constant coupling is obtained after renormalization \cite{Choi:2023cqs, Choi:2023mma}, see (\ref{effcoup}) below. 

We calculate the couplings using the standard perturbation order by order (see e.g. \cite{Peskin:1995ev}). Separating the action to the free $S_0$ and the rest $S_{\text{int}}$ and taking $Z_0[J]=\int {\cal D}\phi e^{-S_0[\phi]}$, we construct
\be
Z[J] = \exp\left( -S_{\text{int}}\left( \frac{\delta}{\delta J}\right)\right) Z_0[J].
\ee
At order $L$, we construct the effective coupling by applying the Feynman rules, see (\ref{genamp}) and Section \ref{sec:contr} below. The corresponding Feynman diagram is characterized by $n$ external scalar fields.
In general, there is more than one diagram contributing to a vertex function 
\be \label{ampsum}
\Gamma_{L}^{(n)} (\p) = \sum_a A_{L}^{(n),a} (\p),
\ee
as we have seen in Fig. \ref{fig:lambda2} for $\Gamma_{2}^{(4)} (p_1,p_2,p_3,p_4)$, for example. Here, $a$ indexes the possible different amplitudes. We can also generalize it to a theory with fields of various spins, in which case the superscript should contain more information on the external fields.

\subsection{The problem of full renormalization}

Now, we wish to compare the theory with the experiment.
We fix the reference coupling at a scale $\p=\underline{\mu^{(n)}}$ by comparing it with the matrix element there
\be \label{phycoup}
 -i\Gamma^{(n)}_{(L)}(\underline{\mu^{(n)}})= i{\cal M}(\text{$n$ scalars})=-i \Gamma^{(n)}_{\text{exp}}+ \O(\hbar^{L+1}),
\ee
generalizing the quartic coupling in (\ref{LambdaRenCond}). In Section \ref{sec:2loop}, we took the reference scale $\mu^{(4)}$ related to $s_0, t_0, u_0$ at which the coupling $\lambda$ is defined. For the mass, we took the reference scale $(\mu^{(2)})^2=m^2$ as the pole mass. Since we expand the theory around these, the scales $s_0, t_0, u_0, m^2$ should not be different too much.

The premise that we can take the renormalization condition (\ref{phycoup}) may be a big assumption. We should guarantee that unknown UV physics beyond our reach must not affect the condition. The Appelquist--Carazzone decoupling theorem \cite{Appelquist:1974tg} and its extension to the scalar mass \cite{Choi:2023cqs, Choi:2023mma} justify it.

We obtain the relation between the bare coupling and the experimental value
\be
\Gamma^{(n)}_{\text{exp}}  = g_B^{(n)} + \sum_{l=1}^L \Gamma^{(n)}_l (\umu).
\ee
We can express the loop-corrected coupling (\ref{Gammaexp}) using the reference one, eliminating the bare one
\be \label{gammasubtraction}
  \Gamma^{(n)}_{(L)} (\p) = \overline \Gamma^{(n)}_{\text{exp}} + \sum_{l=1}^L \left( \Gamma^{(n)}_l (\p) - \Gamma^{(n)}_l (\umu) \right) \stackrel{?}{+} {\cal O}(\hbar^{L+1}),
\ee
in the limit $L \to \infty$. Not only these parameters but the entire effective action can be alternatively parameterized using this physical parameter.

We have seen that just the subtraction worked for the loop corrections to the quartic coupling. However, for the scalar mass, the subtraction was not enough to remove the subleading divergence. 
In general, although the leading divergence in $\Gamma^{(n)}_l (\p)$ is removed in (\ref{gammasubtraction}), we may still have divergences. Let us consider one component $A_L^{(n)}(\p)$ in (\ref{ampsum}) again. This is measured by the superficial degree of divergence $\delta$ in four dimensions of the amplitude $A_L^{(n)}(\p)$ is defined as \cite{Weinberg:1995mt}
\be \label{divdeg}
 \delta = 4 - \sum_f (s_f+1) E_f + \sum_i N_i \Delta_i, 
\ee
where $f$ labels the external legs and $i$ labels the vertices so that it does not depend on the details of the internal part of $\gamma$. Also, $\Delta_i$ is the mass dimension of the corresponding physical parameter, and $E_f,s_f$ are the numbers of external legs and the spin of the field $f$, respectively. In what follows, we may omit the $\gamma$-dependence and simply denote $\delta$ without confusion.

\begin{theorem}[Weinberg \cite{Weinberg:1959nj}] \label{thm:Weinberg}
The degree $\delta$ is the degree of the divergence of the momentum integral if all the momenta go to infinity. Thus, the amplitude is UV finite if $\delta <0$.
\end{theorem} 
That is, the corresponding integrand behaves like $k^\delta$ in the large momentum limit. It is nontrivial to separate the large momenta of the subdiagrams and those of the main diagram. The theorem shows that the large momentum divergence of the overall amplitude is determined by the external legs of the diagram. 
An amplitude formed by renormalizable interactions and involving a loop is divergent if the number of external legs is equal to or smaller than the number of dimensions, 4.

We may impose the same analytic behavior, ``self-similarity'', on the renormalized quantity as on the free particle. This is realized by Taylor expansion and comparison. 
Let $\delta$ be the degree of divergence of the amplitude $A^{(n)}_L$. We have its renormalized coupling \cite{Dyson:1949ha,Zimmermann:1969jj}
\be \label{renormalization} 
 \overline A^{(n)}_L (\p) =\left[1-t^\delta_\umu\right] A^{(n)}_L(\p),
\ee
where $t^\delta_\umu$ is the Taylor expansion operator with the order $\delta$ around the reference momentum $\p = \underline{\mu^{(n)}}$,
\be \label{onshell}
  t^\delta_\umu A^{(n)}_{L} (\p) = \sum_{i=0}^\delta \frac{1}{i!} (p_1^{\alpha_1} - \mu_1^{\alpha_1}) \dots (p_i^{\alpha_i}-\mu_i^{\alpha_i}) \frac{\partial^i  A^{(n)}_{L} }{\partial p_1^{\alpha_1} \dots \partial p_i^{\alpha_i}} (\p = \umu).
\ee
We need $\delta \ge 0$ for renormalizability. The leading term is $A^{(n)}_{L}(\umu)$, similar to the one in (\ref{gammasubtraction}). However, the amplitude $A^{(n)}_L(\p)$ contains divergence if we differentiate $i<L$ times in $p$. In other words, after Tylor expansion, all the coefficients of the degree $\delta$ polynomial are divergent. Although renormalization does not change the total loop-corrected amplitude $A^{(n)}_{(L)}(\p)$, if we only look at a higher-order term, it appears to change as in (\ref{renormalization}). Nevertheless, in the OS scheme, the tree-level value does not change; the effect of the redefinition is hidden in the definition of the bare parameter.

In the BPHZ scheme, we take the zero reference momentum $p^2=0$.
Our renormalized parameters can be transformed from the BPHZ ones by the finite renormalization \cite{Hepp:1966eg} 
\be \label{finiteren} 
\left[1-t^\delta_\umu\right]  A^{(n)}_{L}  =  \left[1-t^\delta_\p\right] A^{(n)}_{L}  + \left[t^\delta_\p-t^\delta_\umu\right]   A^{(n)}_{L}, 
\ee
where all the (three pairs of) terms are finite, having the subtraction structure made finite by Taylor expansions. Since there is no regularization dependence, the only scheme dependence is the reference parameter at different scales. Up to this, any renormalization scheme gives the same result in the {\em effective action.}

The Taylor expansion structure (\ref{renormalization}) originates from the field strength renormalization as in the derivation of the momentum-dependent mass (\ref{SlidingMass}),
\be \label{re-n}
 \phi \equiv \sqrt{Z} \phi_r.
\ee
This is the redefinition of the field, which is, in the $\phi^4$-theory, the only field that should affect all the parts of the amplitude. As a result, the extra terms than $A(\umu)$ in (\ref{renormalization}) take full care of the divergence. After the renormalization, we have ${\cal L}[\phi] = {\cal L_{\text{ren}}}[\phi_r], J = \sqrt{Z} J'$ and
\be 
 Z[J'] = e^{-i E[J']} = \frac{1}{Z[J']_{J'=0}} \int {\cal D}\phi_r \exp \left[ \frac{i}{\hbar} \int d^4x \left( {\cal L_{\text{ren}}}[\phi_r]+J' \phi_r \right) \right]
\ee
We use the strength renormalized field $\phi_r$, and the rest of the perturbation is the same. 

Traditionally, the last terms in (\ref{renormalization}) or the Taylor expanded part have been called counterterms. The counterterm interactions should be regarded as a part of the original Lagrangian, not additionally introduced by {\it ad hoc.} Otherwise, it redefines the theory. What is worse is the newly added interactions may give rise to their loop, which must introduce proliferating counter-counterterms as we have seen in the example in (\ref{disjct}) \cite{Choi:2024cbs}.

Although the field-strength renormalization (\ref{re-n}) must remove all the divergence by counterterms, we do not know such counterterms before calculating the amplitude. In practice, we use the effective action before field-strength renormalization (\ref{effaction}) and obtain counterterms using the renormalization condition. Then, we can remove the remaining divergence using the procedure (\ref{renormalization}). 

Higher order terms in $p_i$'s are given by Taylor expansion coefficients, fixing, for instance, the normalization of the fields that we cannot observe, in view of the examples in (\ref{SlidingMass}) and (\ref{SlidingMass2}). 
Technically, although the operator $t^\delta$ removes the divergence, physically, it means that the behavior of physics should be unchanged in the vicinity of the renormalization point $\p=\underline \mu$ to $\O(|\p-\underline \mu|^n$).
This procedure matches and removes the bare coupling up to $\O(\hbar^{L+1})$, as well as the divergent part.

Having renormalized the main part, the remaining problem is that subamplitudes in $\overline \Gamma^{(n)}_L(\p)$ may contain divergences.
Dyson suggested renormalizing the subamplitude recursively \cite{Dyson:1949ha}. 
After renormalizing the main amplitude, we go down to renormalize a subamplitude. Eventually, this subamplitude may contain another divergent subamplitude again. However, we may repeat and retrack the divergences, and the procedure ceases after finite steps. This top-down program becomes procedural thanks to the Bogoliubov relation and Zimmermann's forest formula, which is included in the BPHZ program \cite{Bogoliubov:1957gp, Hepp:1966eg, Zimmermann:1967}. 

In this paper, we take the opposite, bottom-up route. Namely, we iteratively generate amplitudes using the renormalized subamplitudes. Since a renormalized amplitude can replace a tree-level coupling, we can generate the amplitudes by replacing them with another renormalized amplitude. The renormalization is accomplished if we systematically find the desired amplitude. In fact, we need all the amplitude at a given order as in (\ref{ampsum}) to compare the theory with the experiment, thus generating the complete amplitudes is necessary.

In both approaches, the self-similarity of loop amplitude is important. A subdiagram is no different from the main amplitude in the sense that it and the sub-sub amplitude may have the same structure.

\subsection{Contraction of loops} \label{sec:contr}

With effective couplings, we can generate fully renormalized amplitudes of an arbitrary loop order. To a ``simple'' loop that is renormalized, we may replace a coupling with another renormalized loop amplitude and we may repeat it iteratively. To facilitate this, it is convenient first to define the reverse operation of replacing a loop amplitude with a tree-level coupling.

We have a Feynman diagram $A_L$, whose vertices $\{v_a\}$ and lines (edges) $\{l_{ab}\}$ correspond to the coupling and the propagators, respectively, according to the Feynman rules. Specifying a diagram by a set of lines is convenient because the vertices are automatically selected, while the opposite is not true.
A subdiagram $A_l$ of $A_L$ contains vertices $\{v_\alpha\}$ and lines $\{l_{\alpha \beta}\}$ as subests of $\{v_a\}$ and $\{l_{ab}\}$, respectively. 
If a line connects two vertices of the subdiagram, it also connects those of the whole diagram. 
In this paper, practically every subdiagram is a loop, having vertices along the lines of the loop.

\begin{figure}[t]
\begin{center}
\includegraphics[scale=0.45]{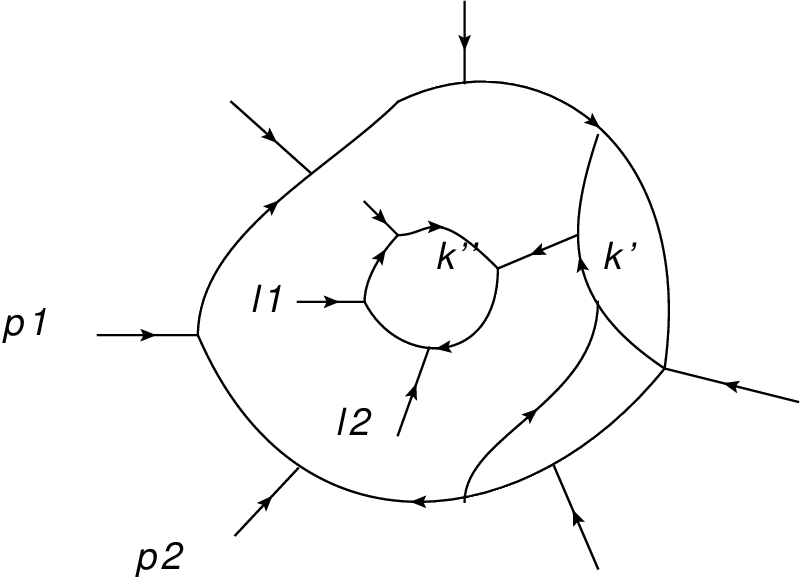} \quad
\includegraphics[scale=0.45]{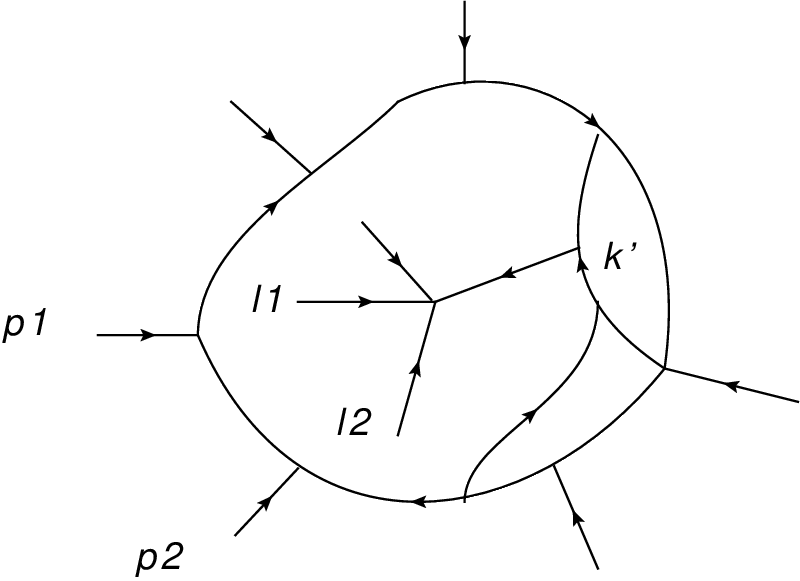}
\end{center}
\caption{A typical multiloop diagram. We can $r$-contract a subgraph $\gamma$ to a point $v$ as long as the resulting tree coupling exists in the Lagrangian. The total inflowing momenta $\{ \ell^v_i \}$ to $\gamma$ and the same to $v$ is conserved.  \label{figcontraction}}
\end{figure}

Without confusion, we denote the amplitude using the same notation as the corresponding Feynman diagram.
A general amplitude in QFT is constructed using the Feynman rules and has the form \cite{Zimmermann:1969jj}
\be \label{genamp}
 A^{(m)}_L (\p) =   \int \frac{d^4 k_1}{(2\pi)^4} \dots  \frac{d^4 k_L}{(2\pi)^4}   \prod_{\{a\}} \big[\overline g(v_a) P_a(\p,\underline{\tilde k_i}) \big]  \prod_{\{l_{ab}\}} \Delta_{ab}(\p,\underline{k_i}).
\ee
Here $\Delta_{ab}(\p,\underline{k_i})$ are propagators along the lines $l_{ab}$. At the vertices $v_a$, we also have constant couplings $\overline g^{(n)}(v_a)$ and polynomials $P_a(\p,\underline{\tilde k_i})$.\footnote{We understand that the product reaches only the next factor, not all the rest factors on the right.} The argument in $\overline g^{(n)}(v_a)$ is for tracking the location in the graph; everywhere they take the same value $\overline g^{(n)}$, which is the renormalized $n$-point couplings $\overline \Gamma_0^{(n)}$ in (\ref{effcoup}) to be obtained later. 

We construct the amplitude using the bare couplings $g_B$ and replace all of them with the renormalized $\overline g$. We do so because the bare couplings are to be corrected (as in the example of the quartic coupling (\ref{lambda1match})). After the SSR renormalization, identifying the small parameters (such as the combination $\lambda^2 [V(s)-V(s_0)]$), the leading order approximation of the corrected coupling is the renormalized coupling $\overline g$ (like $\lambda$ in (\ref{lambda1match})); then, we do not worry about further correction.

We can perform the part of the integrals in (\ref{genamp}) to obtain functions. See Fig. \ref{figcontraction}. To do this, we partition the set of momenta into those of contracted graph $\gamma$ $\{ k''_i \}$ and the rest $\{ k_i \}$
$$
 \{ k_1,k_2,\dots,k_L \} = \{ k_1',k_2',\dots k_{L-l}'\,|\, k''_1 \dots k''_{l} \}, 
$$
with appropriate reordering and renaming. The sub-amplitude is separated such that 
\be \label{replacedint}
 A^{(m)}_L(\p) = \int \frac{d^4 k'_1}{(2\pi)^4} \dots  \frac{d^4 k'_{L-l}}{(2\pi)^4}  A_{l}^{(n)} (\p,\underline{\tilde k'_i}) 
  	 \prod_{\{a\}-\{\alpha \}} \big[\overline g(\cdot) P_{\cdot} (\p,\underline{\tilde k'_i}) \big]  \prod_{\{l_{ab}\}-\{l_{\alpha \beta}\}} \Delta_{\cdot \cdot} (\p,\underline{k'_i}),
\ee
with
\be \label{wgtvcoup}
 A_{l}^{(n)}(\p,\underline{\tilde k'_i}) =\int \frac{d^4 k''_1}{(2\pi)^4} \dots  \frac{d^4 k''_l}{(2\pi)^4}   
 	\prod_{\{\alpha\}} \big[\overline g(v_\alpha) P_\alpha(\p,\underline{\tilde k'_i}) \big]  \prod_{\{l_{\alpha \beta}\}}  \Delta_{\alpha \beta} (\p,\underline{k''_i}),
\ee
such that there is no $k''_i$ dependent functions in the integrand in (\ref{replacedint}) any more. Then $A_{l}^{(n)}(\p,\underline{k'_i})$ is a function of the external and the rest of the momenta. 

The subamplitude depends on the inflowing external momenta $\ell_i^a$ attached to the loop $\gamma$ through the vertices $v_a$, which are linear combinations of the external momenta $\{ p_i \}$ and the subset $\{\tilde k'_i\} \subset \{k_i' \}$.\footnote{There is the canonical way to relate the momenta, called substitution operation \cite{Zimmermann:1969jj}; however, we do not need it.}  Its sum is conserved, like in (\ref{momconserv}),
\be \label{loopconserv}
 \sum_{i=1}^{n} \ell^a_i = 0.
\ee
At each vertex, there are inflowing external momenta $\ell_i^a$ to $v_a$, also summing up to be zero. Especially when collecting the vertices attached to the external legs, the total momenta flowing into all the vertices vanishes because the total momenta inside the loop is zero. After the cancellation, we are left with the external momenta. Thus, we have  (\ref{loopconserv}). This should be, to make sense of (\ref{Gammaexp}).

From the structure of the vertex functions on the same momentum conservation, we can exchange $A^{(n)}$ with any of $\Gamma_j^{(n)}$. 
\begin{definition}[$r$-contraction]
We define {\em $r$-contraction} of an amplitude $A_l^{(n)}(v,\p)$, located at $v$, as the replacement to the tree-level coupling
\be \label{contraction}
  A_l^{(n)}(v,\p) \to \overline g^{(n)}(v), \quad l >0.
\ee
\end{definition}
In most cases, the amplitude we $r$-contract is a subamplitude in a larger one.
The definition (\ref{contraction}) makes sense if the amplitude $A^{(n)}(\p)$ has the same structure on the momentum conservation. 
In every case except the kinetic term $n=2$, the tree-level coupling is constant. Ultimately, what we need is the reverse of the contraction, from a constant coupling $\overline g^{(n)}$ to the loop amplitude like $A^{(n)}(\p).$ Since we are going to ``resolve'' couplings from the Lagrangian, we need a tree-level coupling. 

On the graph side, the above $r$-contraction of a diagram $A_l$ is to send all the vertices $\{v_\alpha \}$ in $A_l$ to one point $v.$ This defines the locations of the contracted vertex and the uncontracted graph in (\ref{contraction}). We disregard the lines $l_{\alpha \beta}$ in $A_l$. When contracting the loop in the propagator, we also drop the vertex. 
This graphical operation also allows the contraction of a line, but we limit our discussion to loops. The subamplitude $A_l$ we contract is not necessarily a single loop; a multi-loop can be contractable at once. A propagator that is not 1PI can also be contractable in the same way.

It can be that $A_l(v)$ is a subdiagram of a diagram $A_L$. We denote the contraction by
\be 
 A_L  \to A_L/A_l(v).
\ee
In the amplitude, it makes sense as a division $A_L^{(n)}/A^{(n_v)}_l(v)$, if we understand that the constant coupling $\overline g^{(n_v)}$ is multiplied.
The graph contraction does not distinguish $r$-contraction from a usual contraction, for which any of the loops with $n$-external lines are contractible, leaving an $n$-point vertex.

\begin{theorem}[contractibility]
For a renormalizable and generic Lagrangian, an amplitude $A_l^{(n)}$ is $r$-contractabile if $\delta(A_l^{(n)}) \ge 0$, that is {\em renormalizabile.}
\end{theorem}
The $r$-contractability requires the existence of $\overline g^{(n)}$ in the Lagrangian, whose interactions are all renormalizable. The amplitude $A_l^{(n)}$ to be $r$-contracted should be renormalizable, thus we need $\delta \ge 0$. For instance, although we deal with renormalizable theory, we can form a one-loop amplitude with arbitrarily many external fields. For this graph, it is possible to shrink the loop to a point in a similar way as contraction, but the resulting operator becomes non-renormalizable. The corresponding amplitude is finite by default, but we do not have the corresponding coupling at the tree level. 

The $r$-contractability is different from renormalizability if the interaction Lagrangian is non-generic. By genericity, we mean the following.
\begin{definition}[generic Lagrangian] \label{defgen}
 A {\em generic} Lagrangian contains all the operators allowed by symmetry of the theory. 
\end{definition}
In the example of a massless scalar theory, $r$-contraction of two-point amplitude may not be possible even if the whole theory is renormalizable. We may generalize it to amplitudes with tensor structure. In gauge theory, the gauge boson mass is forbidden, and an amplitude $\Pi^{\mu \nu}(\p)$ correcting it should have a form $\propto (p^2 g^{\mu \nu} - p^\mu p^\nu)$ satisfying Ward identity. The amplitude is $r$-contractable because we have a nonzero gauge kinetic term with the same tensor structure.

After the contraction, the amplitude becomes
\be \label{contracteddint}
 A^{(m)}_L(\p) = \int \frac{d^4 k'_1}{(2\pi)^4} \dots  \frac{d^4 k'_{L-l}}{(2\pi)^4}  \overline g^{(n)} (v) 
  	 \prod_{\{a\}-\{\alpha \}} \big[\overline g_{ \cdot} P_{ \cdot} (\p,\underline{\tilde k'_i}) \big]  \prod_{\{l_{ab}\}-\{l_{\alpha \beta}\}} \Delta_{\cdot \cdot} (\p,\underline{k'_i}).
\ee
After the contraction, the conservation (\ref{loopconserv}) is then interpreted as the momentum conservation on the vertex $v$.

\subsection{Irreducible amplitudes}

\begin{figure}[t]
\begin{center}
\includegraphics[scale=0.55]{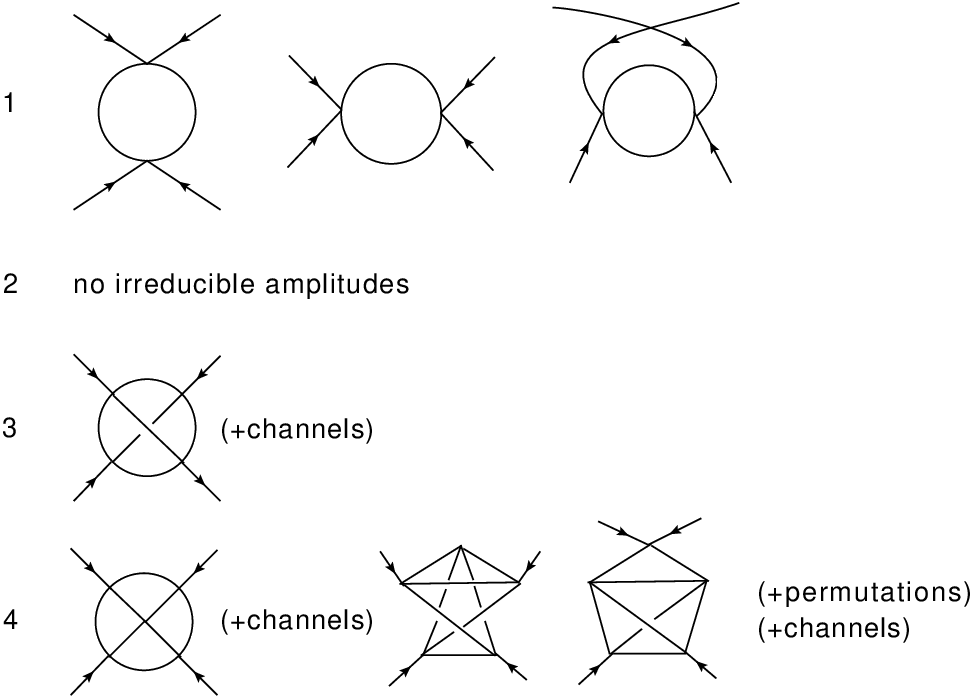}
\end{center}
\caption{Irreducible diagrams up to loop-order four. They are the minimal amplitudes to be renormalized and serve as elementary building blocks of amplitude generations. \label{figirr}}
\end{figure}

In this section, we introduce an irreducible amplitude as the simplest amplitude requiring renormalization.
There are cases where, although the total amplitude is renormalizable, no subdiagram is $r$-contractable. 
We call such an amplitude irreducible.
\begin{definition}[irreducible amplitude] 
We define an {\em irreducible amplitude} as an $r$-contractable one, in which none of the proper subamplitudes can be $r$-contracted. We define every $r$-contractable one-loop amplitude as irreducible. 
\end{definition}
Note that the irreducibility relies on the notion of $r$-contractablity. Starting from an $r$-conrtactable amplitude, $r$-contracting all the {\em subloops} gives us an irreducible diagram. 

One nontrivial example of the irreducible amplitude is a ``Celtic-cross'' amplitude in the $\phi^4$-theory, whose diagram is the last in Fig. \ref{fig:lambda4}, a four-loop diagram with four external legs. It cannot be obtained by resolution by one-loop amplitudes because any loop subamplitude cannot be $r$-contracted. 
Other examples are three-loop amplitudes with no crossing at one vertex of the Cletic-cross amplitude, which is on the first row in Fig. \ref{fig:lambda4} \cite{Bielas:2013rja}. They are the minimal-loop-order irreducible amplitudes for the quartic coupling. Both have the overall logarithmic divergence $\delta=0$ and are free of subamplitude divergence. 

An irreducible amplitude is characterized by non-renormalizable subamplitudes.
\begin{theorem}[irreducibility] \label{thm:irred}
Any subloop amplitude of irreducible amplitude is finite. 
\end{theorem}
This follows from the definition of irreducibility and the Weinberg theorem. Any subamplitude $\gamma$ is non-renormalizable in a renormalizable theory so that we have $\delta(\gamma)<0$. The corresponding amplitude scales as $k^{\delta(\gamma)}$ for large $k$ and converges. The irreducible loop is the {\em minimal} renormalizable amplitude without any subdiagram divergence. 

An irreducible amplitude is renormalized by applying the Taylor expansion operator in (\ref{renormalization}). 
\be \label{compren}
 \overline A^{(n),a_l}_l (\p) = \big[ 1-t_{\mu^{(n)}}^{\delta} \big] A^{(n),a_l}_l (\p),
\ee
with the well-defined degree of divergence $\delta(A^{(n),a_l}_l )$ as in (\ref{divdeg}). Since it has no divergent subamplitudes, the single procedure (\ref{compren}) makes it finite.

Since a number of different diagrams have the same contracted diagram, the contraction is many to one and the resolution is one to many. If one of such can resolve the vertex, others and hence the sum can. 
It is useful to define the sum of all the irreducible amplitudes.
\begin{definition}[irreducible effective coupling] \label{def:reducible} 
We define an {\em irreducible effective coupling} at order $l$ as the complete sum of renormalized irreducible amplitudes  
\be \label{effcomp}
\overline \Gamma_{l}^{(n),\irr} (\p) \equiv \sum_{a_l} \sum_{\text{channels}} \overline A^{(n),a_l}_l (\p).
\ee
Here, the index $a_l$ labels the different types of irreducible amplitudes at the loop-order $L$. The channels are parametrized by the combination of the momenta $\p$. If there is no irreducible amplitude at order $l$, $\overline  \Gamma_{l}^{(n),\irr}(\p)=0$. 
\end{definition}

\begin{figure}[t]
\begin{center}
\includegraphics[scale=0.45]{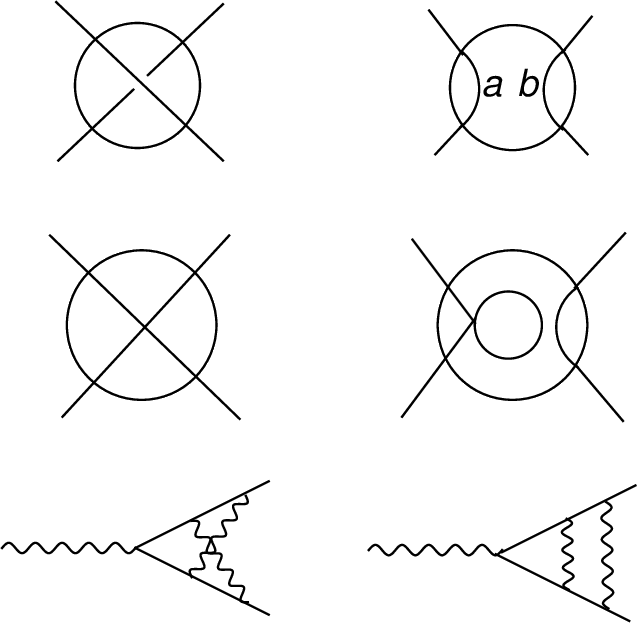}
\end{center}
\caption{Irreducible diagrams are characterized by an internal line connecting non-adjacent vertices. We can generate such ones (left) by flop change, namely, exchanging some vertices of diagrams of effective couplings (right). This process preserves the loop order. \label{figflops}}
\end{figure}

The order of an irreducible amplitude can be arbitrarily large, and we have no general method of generating it. In most cases, nevertheless, we can find ones according to the following observations.
An irreducible loop diagram is characterized by a subloop containing a line that connects {\em non-adjacent vertices}, which makes {\em overlapping loops.} We say two vertices are {\em adjacent} if there is no intermediate vertex on {\em all} the line connecting two vertices.
 This teaches us one helpful way to make irreducible diagrams. We can make one from a reducible loop diagram by rewiring the internal line.

Another property of flop change is that it preserves the number of loops.
\begin{theorem}[flop change]
In generic theory, irreducible diagrams are obtained by flop changes, exchanging the ends of two different internal lines.
\end{theorem}
By definition, it is an action of changing the ending vertices of lines. It does not modify the numbers of vertices and lines, respectively. From the Euler characteristic (\ref{Euler}), it leaves the number of loops $L$ invariant. This means that the only possible variations of the $L$-loop diagram are the flop changes.

Examples are shown in Fig. \ref{figflops}, in which the left on each line is obtained from the right, which can be generated from tree-level coupling. For a single $\phi^4$-theory or QED, the left diagrams are obtained by the flop changes of the right ones, respectively. All the right diagrams can be contracted to quartic coupling or photon-electron interaction. Therefore, it is a good search strategy first to generate an $L$-loop reducible diagram by resolutions and perform the flop change. 

However, not all the irreducible diagrams obtained by flop change. In a non-generic theory with a limited number of couplings, we cannot perform a flop change from an irreducible diagram to a reducible one. For instance, in the first line in Fig. \ref{figflops}, the irreducible diagram on the left cannot be obtained from the reducible diagram if the internal fields are different $a\ne b$. One good remedy is to make the Lagrangian generic by filling in missing operators, then make the diagram and perform a flop change. We can then check whether the irreducible diagrams obtained exist in the original theory. 

Finally, complex irreducible diagrams are obtained from the so-called geographic network at any loop order. Start with a $(L+1)$-polygon and connect all the next-to-adjacent (second) vertices. The resulting diagram has $4 \times (L+1)/2 - 2 = 2L+2$ lines. If we cut two lines to make them external, we may obtain a 4-point vertex with $2L$ lines satisfying the Euler characteristic (\ref{Euler}).

\section{Generating amplitudes using self-similarity}

The above findings on the recursive structure of loop diagrams enable us to synthesize general loop amplitudes. We generalize the results in Section \ref{sec:2loop}, where we have constructed all the quartic and quadratic couplings to two-loop order from one-loop amplitude by resolution. 

\subsection{Resolution and weighted Feynman diagram}

We are ready to construct high-order amplitudes systematically. This can be done by some reverse process of $r$-contraction (\ref{contraction}) accompanied by renormalization. 
\begin{definition}[resolution]
We define {\em resolution} of $\overline g^{(n)}(v)$, located at a vertex $v$, by an amplitude $A^{(n)}(v,\p)$ as the replacement of the tree-level coupling $\overline g^{(n)}(v)$, with  $A^{(n)}(v,\p)$
\be \label{resolution}
  \overline g^{(n)}(v) \to A^{(n)}(v,\p).
\ee
We do the same to the propagator.
\end{definition}
If the location $v_a$ and/or the momentum dependence are obvious, we may omit them.

What we have in mind is the resolution of tree-level couplings to the fully renormalized effective couplings, $\overline \Gamma^{(n)}_0(v) \to \overline \Gamma^{(n)}_L (v,\p), L\ge 1.$ 
The resolution is possible if $A^{(n)}(\p)$ is $r$-contractable to $\overline g^{(n)}$. Also, we may have more than one ``channel'' amplitude having the same external lines, specified by the irreducible diagrams in $A^{(n)}(\p)$. 

On the graph side, a resolution at the vertex $v$ in a graph $\Gamma/A(v)$ by the graph $A(v)$ means we replace a $v$ with $A(v)$. 
The amplitude we have generated by resolution as above {\em increases} the loop-order by $l$, the loop-order of $A^{(n)}.$

\begin{figure}[t!]
\includegraphics[scale=0.9]{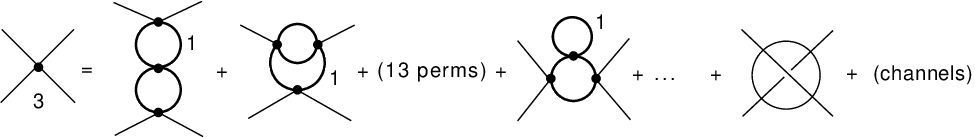}
\vskip 0.5cm
\includegraphics[scale=0.95]{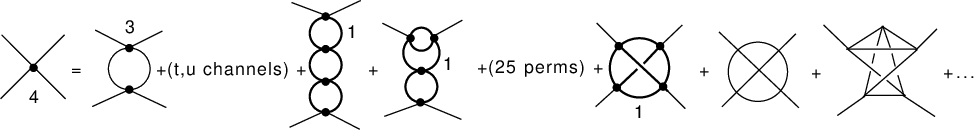}
\caption{The quartic couplings at three-loop and four-loop. Reducible amplitudes of loop-order $l$ are obtained by calculating $m$-loop integrals using loop-order $(l-m)$ couplings and propagators which are denoted by big dots and thick lines, respectively. We resolve it once by an irreducible diagram at one place. There are, in general, irreducible diagrams at high loop order.  \label{fig:lambda4}}
\end{figure}

It is useful not to draw the full diagram $\Gamma$ after the resolution. We may pretend that $A(v)$ is still a constant coupling and also use the contracted diagram $\Gamma/A(v)$. To distinguish this from the actual contracted diagram, we indicate the corresponding vertex using a big dot and indicate the weight in the vicinity as in Fig. \ref{fig:lambda4}.\footnote{Without weights, it is known as a reduced Feynman diagram \cite{Zimmermann:1969jj}.} Likewise, we keep the order $l$ of the propagator $\Delta_l$. 

\begin{definition}[weighted Feynman diagram]
Without $r$-contracting actual amplitude, only on the diagram side, we may contract subdiagrams $\{ \gamma_l \}$ and regard them as tree-level couplings of weights $\{ l\}$.   This replacement gives a weighted Feynman diagram $\Gamma/\{ \gamma_l \}$. We denote the orders $\{ l\}$ in the vicinity of the resulting vertices. We do the same for the replaced lines.
\end{definition}
The contracted subdiagrams are not necessarily irreducible. The original amplitude is unchanged. However, we may formally regard the amplitude (\ref{replacedint}) as a weighted {\em $L-l$-loop} graph.
We have seen many examples in Figs \ref{fig:M1}, \ref{figM2}, \ref{fig:lambda2}, \ref{fig:lambda4}.
To distinguish it, we may write the loop order as the left subscript, like ${}_{L-l} A_L$.

Obviously, since we actually use the loop order $v_i$ amplitude, the total order of an amplitude related to a (sub)diagram $\gamma$ is invariant in the following sense.
\begin{lemma}[loop order]
For a graph $\gamma$ having $m$-loop structure containing the order $l_i$ lines and order $v_i$ vertices. Then
\be \label{bigK}
  \ord_\hbar(A) = m+ \sum l_i + \sum v_i
\ee 
is an invariant after contraction followed by the resolution of the same order.
\end{lemma}

\subsection{Generating high loop-order couplings}

An elementary unit of an additional loop is an irreducible effective coupling in Definition \ref{def:reducible}.
The use of renormalized irreducible amplitudes guarantees divergence-free amplitudes at any loop order.

By successive resolutions, we can generate higher-order amplitudes. In most cases, we can increase the loop order one by one, which dominates the resolution of loop amplitudes. However, this is not always the case because there are irreducible diagrams at various loop orders.
For example, a loop-order 10 amplitude can be made in various ways: (i) An order 9 resolved by an order one irreducible. (ii)  An order 6 resolved by an order 4 irreducible. (iii)  An order 10 irreducible, and so on. We should not consider an order 8 amplitude resolved by order one at two different points because it is included in Case (i). 

We are interested in the total vertex function $\overline \Gamma_{L}^{(n)} (\p)$ that is renormalized, as in (\ref{ampsum}). It is the sum of the complete order-$L$ amplitudes with the same external legs. The accumulated sum $\overline \Gamma_{(L)}^{(n)} (\p)$ can be compared with experiments. We obtain it by inductive resolutions whose total loop order is $L$ in the sense of (\ref{bigK}).

\begin{theorem}[generation] \label{thmgen}
The following procedure {\em generates} the complete renormalizable amplitudes of the loop-order $L$ 
\be \label{generation}
 \overline \Gamma_{(L)}^{(n)}(\p) = \overline \Gamma_{0}^{(n)}+ \sum_{l=1}^L \overline  \Gamma_{l}^{(n)}(\p) , \quad L > 0,
\ee
that are {\em fully renormalized} and the {\em functions of the external momenta $\p$.}
\end{theorem}
\begin{enumerate}
\item Start with $n$-point couplings $\overline \Gamma_0^{(n)}$, which are to be determined by recursively. The pole mass determines a propagator $= \overline \Delta^{-1}_0(p^2) \equiv \overline \Gamma_0^{(2)} = p^2 - m^2$. In the OS renormalization, by definition $\overline \Gamma^{(n)}_{\text{exp}} = \overline \Gamma^{(n)}_{0}$.

\item Assume we have all the finite $n$-point couplings $\overline \Gamma_k^{(n)},$ as below, of order $k=0,\dots l-1$, for all $n$. In practice, we may prepare only the necessary ones.
\item We obtain the order-$l$ reducible amplitudes from those of $m$-loop $\overline \Gamma^{(n)}_{m} (\p), m<l,$ by resolving vertices with the effective couplings $\overline g^{(n_a)} \to \overline \Gamma_{l-m}^{(n_a),\irr}$ in (\ref{effcomp}), {\em one by one,} 
\be \label{recform}
 {}_m \overline \Gamma^{(n)}_l (\p) =  \sum_{v_a,l_{ab}}  S_{v,l}^{-1} \overline  \Gamma^{(n)}_{m} (\p) \Big|^1_{ g^{(n_a)} \to \overline \Gamma_{l-m}^{(n_a),\irr}} , 
\ee
where $v_a$ and $l_{ab}$ are vertices and lines of each component of $\overline \Gamma^{(n)}_{m} (\p)$, respectively. $\overline g^{(n_a)}$ are the coupling $\overline g^{(n)}$ at the vertices $v_a$. Similarly, the propagator is modified by $\overline g^{(2_{ab})} \to \overline \Gamma_{l-m}^{(2_{ab}),\irr}.$ See below for the others.

\item
The replacement is done component by component, defined in (\ref{effcomp}), as follows. 
\begin{itemize}
\item If the subamplitude $A_{l-m}$ is disjoint or overlapping from that of $A_l$, the operation applies to the origianal $m$-loop amplitude $A_l/A_{l-m}$ only.
\item If the $A_{l-m}$ is nested in $A_l$, the operation applies to the total amplitude $A_l$.
\end{itemize}
A renormalized amplitude $\overline A^{(n)}_{l-m}$ and hence the irreducible effective coupling $\overline \Gamma_{L-m}^{(n),\irr}$ in (\ref{effcomp}) may replace the above subamplitude $A_{l-m}$, which is finite. 
For multiply nested or disjoint amplitude, we can do the procedure one by one and the condition is well-defined.

\item Modify the symmetry factors from the symmetry of the resulting diagram, if necessary. For each channel in $\Gamma_{l-m}^{(n)}$, decomposed as in (\ref{wgtsum}), we obtain different $A^a_l$ from $A^a_{l-m}$, so we multiply the combinatoric factor
\be
 S_{v_a} =  \frac{\text{(the order of enhanced symmetry in $A^a_l$)}}{\text{(the order of reduced symmetry in $A^a_{l-m}$)}}.
\ee
We may define the same for the propagators on the lines $ S_{l_{ab}}$.

\item The total sum of the above amplitudes gives the vertex function
\be \label{wgtsum}
 \overline \Gamma^{(n)}_l (\p) = \sum_{m=0}^{l-1}  {}_m \overline  \Gamma_l^{(n)}  (\p) 
\ee
is the order-$L$ coupling that contains no divergence, including those of subdiagrams. 

\item
Iterate the above procedure until we arrive at the desired order $L$. Collecting all, we obtain the complete sum (\ref{generation}).
\end{enumerate}
Proof of completeness: The method is iterative, and the generated diagrams are exhaustive. Using the base as an order-$m$ amplitude, we resolve one of the vertex by an irreducible amplitude. In $\phi^4$-theory, one-loop vertex function $\overline \Gamma^{(4)}_1(\p)$ is obtained by resolution of the constant coupling $\overline g^{(4)}$ by the order-one irreducible one $\overline \Gamma^{(4),\irr}_1(\p)=\overline \Gamma^{(4)}_1(\p).$ Since irreducible amplitudes contain no further $r$-contractable subdiagrams, there is no intermediate amplitude between the two. After the inductive construction of reducible amplitudes, the rest of the order-$l$ amplitudes are all irreducible by definition. Note that although any component of $\overline \Gamma^{(n)}_l$ can become a coupling, a reducible one gives rise to the overproduction of amplitudes.

Proof of finiteness:  At each step, the original amplitude $\overline \Gamma_l/\overline \Gamma_{l-m}$ with tree-level coupling is finite by induction. The irreducible amplitude $\overline \Gamma_{l-m},m>0,$ resolving the coupling in $\overline \Gamma_l/\overline \Gamma_{l-m}$ is finite by construction. The only amplitude to be renormalized is the new irreducible coupling $\overline \Gamma_{l-m}(\p)$. The generation process is inductive and guarantees that all amplitudes are finite.

The OS regularization is defined by the renormalization conditions $\Gamma^{(n)}_{l} (\underline{\mu^{(n)}}) = 0$ for all $l>0$. In this case, we always have $\overline \Gamma^{(n)}_{0} = \overline \Gamma^{(n)}_{\text{exp}}$ and do not need the matching procedure (\ref{effcoup}).
In other schemes, the fully renormalized vertex function $\overline \Gamma^{(n)}_{0}$ is matched by the equation
\be \label{effcoup}
 \overline \Gamma^{(n)}_{\text{exp}} = \overline \Gamma^{(n)}_{0}  + \sum_{l=1}^L  \overline \Gamma^{(n)}_{l} (\underline{\mu^{(n)}}) + {\cal O}(\hbar^{L+1}).
\ee
In other words, regarding $\overline \Gamma^{(n)}_{l}(\p)$ as functions of $\overline \Gamma^{(n)}_{0}$, $\overline \Gamma^{(n)}_{0}$ is the solution to the recursion equation (\ref{effcoup}).
The $\overline \Gamma^{(n)}_{\text{exp}}$ is the observed value at the defining scale $\p = \underline{\mu^{(n)}}$. 
As in (\ref{effcoup}), the parenthesis in the subscript means the corresponding quantity is the accumulated quantities up to the denoted order.

After loop corrections and renormalization, a physical parameter $\overline \Gamma_L^{(n)}(\p)$ acquires the momentum dependence. It is important to note that, due to the subtraction structure (\ref{compren}), the whole effective action, and hence the couplings $
\overline \Gamma_L^{(n)}(\p)$ are finite, independent of regularization \cite{Zimmermann:1969jj}, and parametrically small \cite{Choi:2023cqs, Choi:2023mma}. Although we have not calculated higher-order corrections, they are there to make the perturbation small. Unrenormalized amplitudes $\Gamma_{L}^{(n)}(\p)$ do not have this feature, so they cannot play the role of effective coupling in perturbation.

We define also weight $L$ propagator using the renormalized mass $m_{(L)}(\p)$ inside the propagator
\be \label{propagator}
 \overline \Delta_{(L)}(p^2) = \frac{i}{\overline \Gamma^{(2)}_{(L)}(\p)}, 
\ee 
which generalizes the scalar mass (\ref{SlidingMass2}).
What if we have the following connection of two different 1-particle-reducible (1PR) amplitudes? Consider the amplitude of the schematic form
\be \label{two1pr}
 \Delta_{(l)} \tilde\Sigma_a \Delta_{(l)} \tilde\Sigma_b \Delta_{(l)},
\ee
where $D_{(l)}$ is a propagator and $\tilde\Sigma_i$ is a self-energy.
We may use an 1PI amplitude $\Sigma_c$ containing both $\tilde\Sigma_a$ and $\tilde\Sigma_b$. Its replacement
\be \label{LCMrep}
\Delta_{(l)} \tilde\Sigma_c \Delta_{(l)} \tilde\Sigma_c \Delta_{(l)}
\ee
contains the original one (\ref{two1pr}), giving the correction
\be
 m_{(l)}^2 (p^2) \to m_{(l)}^2(p^2) + \tilde \Sigma_c(p^2) .
\ee
We may think of a common multiple amplitudes.
One way to make the common multiple amplitudes is to replace one propagator of $\Sigma_a$ to contain $\Sigma_b$, or vice versa. 

Theorem \ref{thmgen} is stated for a theory with a scalar. However, it is not difficult to generalize to one with more fields with different spins since the generic amplitude has a similar structure. 

Summarizing, we have calculated the perturbative effective action up to order $L$ that is fully renormalized. The above generating method deals with amplitudes that have no divergences. 
Since the vertex function is nothing but a 1PI correlation function in QFT, we have the following theorem.
\begin{theorem}[finiteness of renormalizable QFT] \label{thm:noinf}
In a renormalizable QFT, although the bare couplings in the Lagrangian or the individual loop corrections to them suffer conceptual problems related to infinity, no infinity appears in the renormalized effective action.
\end{theorem}
For instance, each term on the RHS in Eq. (\ref{lambdacorr}) seems not finite, the entire combination is. This argument applies to a general theory when the amplitudes in the effective action take the form (\ref{genamp}).

\subsection{Examples}

In $\phi^4$-theory, a general reducible order-$3$ quartic coupling is obtained as a weighted two-loop amplitude. Since we do not have order-two irreducible amplitudes, there are no weighted one-loops. 

First, we recall the loop-order two amplitudes in (\ref{Gamma42})
\begin{align}
 6 i   \overline \Gamma^{(4)}_{2}(\p) &= (-6i\lambda)^3 \big( \overline W_s(s) +  \overline W_t(t) +  \overline W_u(u) \big),  \\
  \overline W_s(s) &=\overline W_8 (s) +\overline  W_\circledcirc (s), 
\end{align}
and similar for $t$ and $u$ channels. We have 27 terms in total. Let us first focus on the three amplitudes in $\overline W_8 (s)$
\be \label{W8resol}
\begin{split}
 i \overline W_8 (s)=  \int &\frac{d^4k_1}{(2\pi)^4}\frac{d^4k_2}{(2\pi)^4}
\left[ D_0((p_3+p_4+k_1)^2)D_0(k_1^2) -D_0((p_3+p_4+k_1)^2)D_0(k_1^2)\big|_{s=s_0} \right]  \\
&\quad \times \left[ D_0((p_1+p_2+k_2)^2)D_0(k_2^2) -D_0((p_1+p_2+k_2)^2)D_0(k_2^2)\big|_{s=s_0} \right]
\end{split}
\ee
There are couplings $\lambda$ at the three vertices $v_a,a=1,2,3$, with different conserved momenta, which can be replaced with the following effective coupling
\begin{align}
\overline  \Gamma_1^{(4_1),\irr}(p) & = \overline \lambda_1(-p_3-p_4-k_1,k_1,p_4,p_3), \\
\overline  \Gamma_1^{(4_2),\irr}(p) & = \overline \lambda_1 (-k_2,p_1+p_2+k_2,-k_1,p_3+p_4,k_1), \label{Gamma1irr} \\
\overline  \Gamma_1^{(4_3),\irr}(p) & = \overline \lambda_1(p_1,p_2,-p_1-p_2-k_2,k_2).
\end{align}
Here $\lambda_1(p_1,p_2,p_3,p_4)$ is defined in (\ref{lambda1}). For this $\phi^4$-theory we have only one type of coupling, so all the three replacements may give rise to the same amplitude, so we assign the multiplicity $1/3$ on each, giving
\be 
\begin{split}
 \frac{i}{3} & \overline W_8 (s)\Big|^1_{\overline g^{(n_a)} \to \overline \Gamma^{{(n_a)},\irr}_{1} }\\
 & =   \int \frac{d^4k_1}{(2\pi)^4}\frac{d^4k_2}{(2\pi)^4}
\left[ D_0((p_3+p_4+k_1)^2)D_0(k_1^2) -D_0((p_3+p_4+k_1)^2)D_0(k_1^2)\big|_{s=s_0} \right]  \\
&\quad \times  \left[ D_0((p_1+p_2+k_2)^2)D_0(k_2^2) -D_0((p_1+p_2+k_2)^2)D_0(k_2^2)\big|_{s=s_0} \right] \\
& \quad \times \frac{1}{3}  \sum_{a=1}^3 \overline  \Gamma_1^{(4_a),\irr}(p).
\end{split}
\ee
In particular, the contribution from the middle vertex $\Gamma_1^{(4_2),\irr}(p)$ is the function of both loop momenta $k_1,k_2$, so the resulting amplitude becomes
\be
\begin{split}
  -3 i \lambda & \int \frac{d^4k_1}{(2\pi)^4}\frac{d^4k_2}{(2\pi)^4}
\left[ D_0((p_3+p_4+k_1)^2)D_0(k_1^2) -D_0((p_3+p_4+k_1)^2)D_0(k_1^2)\big|_{s=s_0} \right]  \\
&\quad \times  \left[ D_0((p_1+p_2+k_2)^2)D_0(k_2^2) -D_0((p_1+p_2+k_2)^2)D_0(k_2^2)\big|_{s=s_0} \right] \\
 &\times \int \frac{d^4k}{(2\pi)^4}\big[D_0((p_1+p_2+k)^2)D_0(k^2)+  D_0((k_1+k_2+k)^2)D_0(k^2) \nonumber \\
\quad & +  D_0((p_1+p_2+k_2-k_1+k)^2)D_0(k^2)
- (s=s_0)\big] .\label{intvertex} 
\end{split}
\ee
is unfactorized and now has a nested structure. The exception is the first component, which is renormalized as the disjoint component as in the two-loop case in (\ref{disjct}). 

Now consider the amplitude
\be
\begin{split}
 \overline W_\circledcirc (s) & \equiv \frac12 \int \frac{d^4k_1}{(2\pi)^4}\frac{d^4k_2}{(2\pi)^4}D_0((p_3+p_4+k_1)^2)D_0(k_1^2)  \\
 & \qquad \times  \Big(D_0((p_3+k_1+k_2)^2)D_0(k^2) -  \big[D_0((p_3+k_1+k_2)^2)D_0(k_2^2)\big]_{s=s_0}\Big) ,
\end{split}
\ee
whose diagram is less symmetric: we have two equivalent vertices only. Nevertheless, it is remarkable that the same multiplicity factor $1/3$ should be assigned. We replace one coupling with one of the following couplings
\be
\begin{split}
& \overline \lambda_1 (k_1,-k_1-k_2-p_3,k_2,p_3), \\
& \overline \lambda_1 (k_1+k_2+p_3,-k_1+p_3,p_4,-k_2),\\
& \overline \lambda_1 (p_1,p_2,k_1-p_3,k_1).
\end{split}
\ee
One of the amplitudes has the same diagram as two obtained from (\ref{W8resol}), giving the total multiplicity 1.

Likewise, we can resolve the internal lines by resolutions $\overline g^{(2_a)} \to \overline \Gamma^{{(2_a)},\irr}_{1}(\p,\k)$. Every amplitude has $I=4$ lines, and we assign the multiplicity $1/4$ on all of them.

We have one type of irreducible diagram in three channels, as in Fig. \ref{fig:lambda4}. They are all non-planar. In one channel, we have
\be \label{v3irr1}
\begin{split}
 6i \Gamma_{3}^{(4),\irr}(\p)& =(-6i\lambda)^4    \int \frac{d^4k_1}{(2\pi)^4}\frac{d^4k_2}{(2\pi)^4} \frac{d^4k_3}{(2\pi)^4}D_0(k_1^2)D_0((k_1+p_1-k_2)^2)D_0(k_2^2) \\
 & \quad \times D_0((k_2+p_2-k_3)^2)D_0(k_3^2) D_0((k_3-k_2+p_1+p_4)^2)\\
& \quad + \text{ (channels).}
\end{split}
\ee
It is renormalized as usual $\overline \Gamma_{3}^{(4),\irr}(\p) = (1-t^0_\umu) \Gamma_{3}^{(4),\irr}(\p)$.

Collecting, we have the sum of the renormalized three-loop amplitudes
\be \label{v3} 
\begin{split}
i \overline \Gamma^{(4)}_{3}(\p)  &= \frac{1}{3} \sum_{\text{3 vertices}} i \overline  \Gamma^{(4)}_{2}(\p) \Big|^1_{\overline g^{(4_a)} \to \overline \Gamma^{{(4_a)},\irr}_{1} } +\frac{1}{4} \sum_{\text{4 lines}} i  \overline \Gamma^{(4)}_{2}(\p) \Big|^1_{\overline g^{(2_a)} \to \overline \Gamma^{{(2_a)},\irr}_{1} }   + i \overline  \Gamma^{(4),\irr}_3 (\p). 
\end{split}
\ee

Likewise, for the order-4 vertex, we make use of the previous result. The reducible parts (\ref{v3}) are obtained from $\overline  \Gamma^{(4)}_{1}(\p)$ and $\overline  \Gamma^{(4)}_{3}(\p)$  by replacing every vertex and line by irreducible order three and order one coupling, respectively,
\be \begin{split}
i \overline  \Gamma^{(4)}_4(\p)  &= \frac{1}{2}  \sum_{\text{2 vertices}} i\overline  \Gamma^{(4)}_{1}(\p) \Big|^1_{g^{(4_a)} \to\overline  \Gamma^{{(4_a)},\irr}_{3} }  + \frac{1}{4} \sum_{\text{4 vertices}}i \overline  \Gamma^{(4)}_{3}(\p) \Big|^1_{g^{(4_a)} \to\overline  \Gamma^{{(4_a)},\irr}_{1} } \\
& \quad+ \frac{1}{6} \sum_{\text{6 lines}}i\overline  \Gamma^{(4)}_{3}(\p) \Big|^1_{g^{(2_a)} \to \overline  \Gamma^{{(2_a)},\irr}_{1} } +\sum_{\irr_4} i\overline \Gamma^{(4),\irr}_4 (\p).
\end{split}
\ee
We have no order-3 irreducible propagators.
And we have two types of irreducible amplitudes in various channels, contributing to $\overline \Gamma^{(4),\irr}_4 (\p)$.

This is to be applied to the two-point correlation function and the effective operator $\overline \Gamma^{(2)}$. The resulting propagator is also obtained as the inverse $\Delta(\p) = i/\overline \Gamma^{(2)}(\p)$ as in (\ref{propagator}).

\subsection{Overlapping divergence}

The above generating method guarantees no subdiagram divergences. 
The prescription (\ref{renormalization}) also removes the overall divergence. 

\begin{figure}[t]
\begin{center}
\includegraphics[scale=0.75]{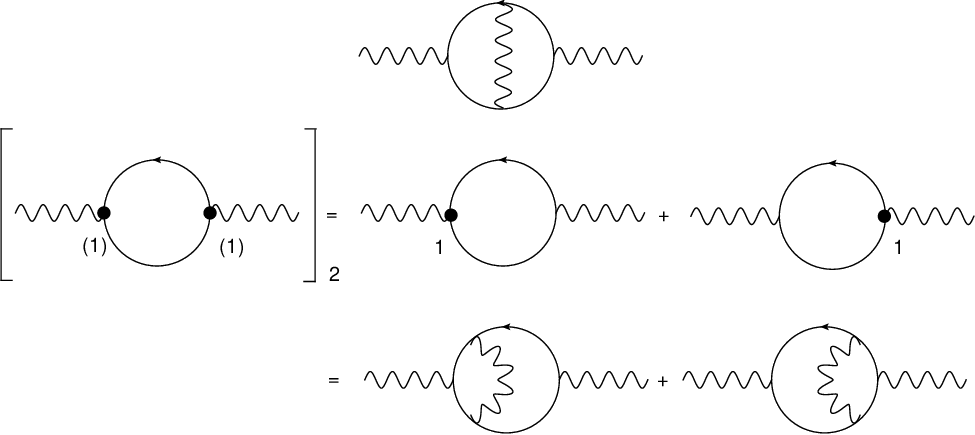}
\end{center}
\caption{Overlapping divergence in a two-loop diagram in QED (top). It is automatically taken care of if we generate it (as the sum of two diagrams) using renormalized couplings $[-i.e.,_{(1)}(\p)]$ in the weighted one-loop vacuum polarization graph (bottom). The last two diagrams give the same amplitudes and are summed up to yield the correct multiplicity. \label{figPi2}}
\end{figure}

This might not hold if we happen to have, as a result of the resolution, {\em overlapping loops}. An overlapping diagram has more than one loop sharing an internal line, like the sunset diagram in Fig. \ref{figM2} and the one in Fig. \ref{figPi2}. It has an ambiguous origin of divergence since the shared propagator carries the momenta of both loops. 

In our iterative renormalization, the divergence from the overlapping loop is automatically taken care of in a symmetrical way. Consider a two-loop vacuum polarization diagram in QED, the top panel in Fig. \ref{figPi2}. We can obtain it by resolving the weight of one vertex of the one-loop Feynman diagram as in the middle panel in Fig. \ref{figPi2}, in which there are two cases. These {\em two cases appear automatically} if we generate the weighted ``one-loop'' amplitude by resolving the photon propagator $\Pi^{\mu \nu}_{2}(p^2)$. 
\be
 -i \Pi^{\mu \nu}_{2}(\fs p)= - \int \frac{d^4k}{(2\pi)^4} \tr (-i e \gamma^\mu(k,p)) S(k) (-i e \gamma^\nu(k,p)) S(k+p) .
\ee
There are two vertices that we may replace with one-loop vertex functions $\overline \gamma^{\mu}_{1}(p^2)$ that are renormalized, finite and momentum-dependent.

Suppose we resolve asymmetrically one of the vertices in the weighted one-loop diagram, as in the lower panel in Fig. \ref{figPi2}. The resolved part becomes the subdiagram, which is divergence-free by construction. We only need to renormalize the overall part. We also have another resolved loop from the other vertex. So, although each is assumed asymmetric, we expect the total amplitude to be. 

Extracting a two-loop amplitude by resolution as in the middle panel in Fig. \ref{figPi2}, we have the unrenormalized part
\be \begin{split}
 -i \Pi^{\mu \nu}_{2*}(p^2)&= - \frac12 \int \frac{d^4k}{(2\pi)^4} \tr[-i e \overline \gamma_1^\mu(k,p+k)] S(k) [-i e \overline \gamma^\nu(p+k,k)] S(p+k) \\
 &\quad  - \frac12 \int \frac{d^4k}{(2\pi)^4} \tr[-i e \overline \gamma^\mu(k,p+k)] S(k) [-i e \overline \gamma_1^\nu(p+k,k)] S(p+k) \\
 &=\frac{ie^4}{4} \int \frac{d^4k}{(2\pi)^4}\frac{d^4k'}{(2\pi)^4}  \frac{1}{(k'-k)^2}   \tr\gamma^\mu\big( S(k'+p) - S(k') \big) \gamma^\rho S(k') \gamma_\rho S(k) \gamma^\nu S(k+p)  \\
 & \quad  + \frac{ie^4}{4} \int \frac{d^4k}{(2\pi)^4}\frac{d^4k'}{(2\pi)^4}  \frac{1}{(k'-k)^2}   \tr\gamma^\mu S(k) \gamma^\nu  \big( S(k'+p) - S(k') \big) \gamma^\rho S(k'+p) \gamma_\rho  S(k+p), 
\end{split}
\ee
where the trace acts on all the gamma matrices and the propagators on the right, using the standard notation \cite{Peskin:1995ev}. 
Although the two terms are the same, we rename the internal momenta in the last line as
\be 
 \frac{ie^4}{4} \int \frac{d^4k}{(2\pi)^4}\frac{d^4k'}{(2\pi)^4}  \frac{1}{(k'-k)^2}   \tr\gamma^\mu S(k') \gamma^\nu  \big( S(k+p) - S(k) \big) \gamma^\rho S(k+p) \gamma_\rho  S(k'+p).
\ee
This makes the two amplitudes have the same momenta at each vertex.

Since we understood the amplitude hierarchical and essentially the divergence comes from the bigger diagram as in the lower panel in Fig. \ref{figPi2}, we renormalize the total amplitude
\be \label{QEDoverlap} \begin{split}
 \overline \Pi^{\mu \nu}_2(p^2) &=  (1-t^2_0) \Pi^{\mu \nu}_2(p^2) \\
  &=\frac{e^4}{4} \int \frac{d^4k}{(2\pi)^4} \frac{d^4k'}{(2\pi)^4}\frac{1}{(k'-k)^2}   \tr\gamma^\mu  \big( S(k'+p)  -  S(k') \big) \gamma^\rho S(k') \gamma_\rho  S(k) \gamma^\nu S(k+p) \\
  &\ + \frac{e^4}{4} \int \frac{d^4k}{(2\pi)^4}\frac{d^4k'}{(2\pi)^4}  \frac{1}{(k'-k)^2}   \tr\gamma^\mu S(k') \gamma^\nu  \big( S(k+p) - S(k) \big) \gamma^\rho S(k+p) \gamma_\rho  S(k'+p) \\
&\ - \frac{e^4}{4} p^2 \int \frac{d^4k}{(2\pi)^4}\frac{d^4k'}{(2\pi)^4} \frac{1}{(k'-k)^2}  \tr\gamma^\mu  \frac{dS}{d(p^2)}(k')\gamma^\rho S(k') \gamma_\rho  S(k) \gamma^\nu S(k)\\
 &\ - \frac{e^4}{4} p^2 \int \frac{d^4k}{(2\pi)^4}\frac{d^4k'}{(2\pi)^4} \frac{1}{(k'-k)^2}   \tr\gamma^\mu    
 S(k') \gamma^\nu \frac{dS}{d(p^2)}(k) \gamma^\rho  S(k) \gamma_\rho S(k').
\end{split}
\ee
Note that $\Pi^{\mu \nu}_2(0)=0.$ 
Now, we may regard (\ref{QEDoverlap}) as the Taylor expansion of the following
\be \label{overlap} 
\begin{split}
 \overline \Pi^{\mu \nu}_2(p^2) &\simeq \frac{e^4}{4}\int \frac{d^4k}{(2\pi)^4} \frac{d^4k'}{(2\pi)^4} \tr\gamma^\mu   \frac{1}{(k'-k)^2}  \big( S(k'+p) -  S(k')  \big) \gamma^\rho S(k') \gamma_\rho S(k) \gamma^\nu \big (S(k+p)-S(k) \big)
\end{split}
\ee
We obtained the amplitude, which is {\em symmetric} under the exchange of the two loops.
In fact, the two amplitudes are the same as in the top panel since they are related by 180-degree rotation accompanied by $p \to -p$.

Moreover, the form (\ref{overlap}) looks like we have renormalized each one-loop separately, neglecting the common propagator. This proved the following theorem in the special case.
\begin{theorem}[Berg\`ere and Zuber \cite{Bergere:1974zh}] \label{thm:bz}
In overlapping two loops, divergence comes from each loop separately, not from the whole amplitude. 
\end{theorem}
Thus, for the subamplitudes in the overlapping amplitude, we treat them like disjoint ones in renormalization.

When we resolve a weighted vertex in the loop, we may generate an overlapping loop diagram. In our case, and the ones seen in Section \ref{sec:2loop}, the overlapping arises along the time-ordering of the external legs from the corresponding Green function. In this case, the resolution gives rise to more than one amplitude, which automatically gives the correct combinatorics.  We can mechanically apply the resolution procedure. On the other hand, as in the case of the sunset and $W_8$ amplitudes, if the overlapping arises transverse to the time-ordering, we need to modify the symmetry factor.

\section{Renormalization using weighted Feynman diagrams}

So far, we have focused on generating amplitudes $\overline \Gamma_L^{(n)}$ that are fully renormalized. Traditionally, the problem was the reverse: how do we renormalize {\em a given} unrenormalized amplitude $A_L$? We only need to {\em find} the renormalized one $\overline A$ among the components $\overline \Gamma_L^{(n)}$. We develop a method of renormalization making use of it.

\subsection{The forest formula}

We have done the bottom-up renormalization for a general amplitude. However, it is useful to compare it with the traditional one, the top-down method, using the recursion relation (\ref{BP}). 

The Bogoliubov recursion relation is the general procedure for full renormalization \cite{Bogoliubov:1957gp}. 
\begin{theorem}[Bogoliubov and Parasuik \cite{Bogoliubov:1957gp}, Hepp \cite{Hepp:1966eg}]
A given amplitude $A_\Gamma$ corresponding to a graph $\Gamma$ is renormalized by
\be \label{BP}
\overline A_\Gamma = (1-t^\Gamma) A'_\Gamma,
\ee
where $A'$ is recursively defined as
\be \label{BPrec}
 A'_\gamma =  A_\gamma + \sum_{\{\gamma_1,\dots \gamma_c\}} A_{\gamma/\{ \gamma_1\dots \gamma_c \}} \prod_{i=1}^c (-t^{\gamma_i}) A_{\gamma_i},
\ee 
and $\{\gamma_1,\dots,\gamma_c\}$ is a set of mutually disjoint connected subdiagrams of $\gamma$.
\end{theorem}
Zimmermann found the set of ``forests'' below solves the recursion relation (\ref{BPrec}) \cite{Zimmermann:1969jj}. 
\begin{theorem}[Zimmermann \cite{Zimmermann:1969jj}] For the amplitude $A$ with the graph $\Gamma$, we obtain the renormalization $\overline A$ by 
\be \label{forestformula}
 \overline A = (1-t^\Gamma) \sum_{U \in {\cal F}} \bigg[ \prod_{\gamma \in U} (-t^\gamma) \bigg] A,
\ee
where the forests $U$ of the $\Gamma$ and their family ${\cal F}$ are to be defined below.
\end{theorem}

Since our generated amplitude $\overline \Gamma^{(n)}_L$ is fully renormalized, we may extract the forest solving the relation (\ref{forestformula}). 
\begin{definition}[forests]
Let the renormalization of an amplitude $A^{(n)}_L$ be $\overline A^{(n)}_L$. This $\overline A^{(n)}_L$ contains the Taylor expansion of the subamplitudes $\gamma$, that is, a number of terms acted by the product of Taylor operators $\prod t^\gamma$. For each term, we contract the corresponding subgraph $\gamma$ when applying $t^\gamma$. We define the {\em forest} $U$ as the set containing the set of graphs undergoing the contractions simultaneously by $\prod t^\gamma$.
\end{definition}

Since we are only interested in the divergence structure, we abbreviate a general amplitude (\ref{genamp}) with the corresponding Feynman graph $\gamma$ as
\be
 A = \int_{\gamma} I_\gamma,\quad I_\gamma = \Delta_\gamma g_\gamma,
\ee
with the total product of the propagators $\Delta_\gamma$ from all the lines in $\gamma$ and the total product of the couplings $g_\gamma$ from all the vertices in $\gamma$.\footnote{The integral can be omitted in our discussion.} We suppressed the momentum dependence.

First, we have nesting diagrams, defining the subgraph. Consider a diagram $\gamma$ and its subdiagram $\gamma_i \subset \gamma$. The amplitude is
\be
 A = \int_{\gamma/\gamma_i} I_{\gamma/\gamma_i}  \int_{\gamma_i} I_{\gamma_i}
\ee
Further contractions are possible for further nesting structures.
Performing subamplitude renormalization on $\gamma_i$, we have
\be
 \int_{\gamma/\gamma_i} I_{\gamma/\gamma_i}  \int_{\gamma_i} [1-t^{\gamma_i}] I_{\gamma_i} 
\ee
The full renormalization of the amplitude is
\be \begin{split}
 \overline A &= [1-t^\gamma] \int_{\gamma/\gamma_i} I_{\gamma/\gamma_i} \int_{\gamma_i} [1-t^{\gamma_i}] I_{\gamma_i}  \\
  & = \int_{\gamma/\gamma_i} I_{\gamma/\gamma_i}  \int_{\gamma_i} I_{\gamma_i}   - t^{\gamma_i} \int_{\gamma/\gamma_i} I_{\gamma/\gamma_i} \int_{\gamma_i}  I_{\gamma_i}-t^\gamma \int_{\gamma/\gamma_i} I_{\gamma/\gamma_i}  \int_{\gamma_i} I_{\gamma_i} 
  +t^\gamma t^{\gamma_i} \int_{\gamma/\gamma_i} I_{\gamma/\gamma_i} \int_{\gamma_i}  I_{\gamma_i} .
\end{split}
\ee
The first term in the second line is the original $A$.
Note that the $t$-operator on the right acts first.
Now, contract the part of the graph if it is acted by the Taylor expansion operator $t$. The $t$-acted part is the counterterm and we may indicate it in the diagram. Reading from the amplitude gives the forests\footnote{Thus, we have four forests. We omit their labels.}
\be
 U_{\text{nes}} =  \{ \gamma, \gamma_i\}, \{\gamma \}, \{\gamma_i \}, \emptyset . 
\ee
The inductive method of generating higher loop diagrams using effective couplings corresponds to considering nested forests of subdiagrams. If a coupling belongs to a higher-order coupling, the corresponding lower-order subdiagram is a part of the higher-order (sub)diagram. We say that the former is {\em nested} in the latter.

Next, we consider the product amplitude from disjoint diagrams $\gamma_i, \gamma_j, \gamma_i \cap \gamma_j$.
We have
\be
 A =  \int_{\gamma_i} I_{\gamma_i}   \int_{\gamma_j} I_{\gamma_j} .
\ee
They make independent coupling when contracted so that they can be separately renormalized
\be
\begin{split}
\overline A &= [1-t^{\gamma_i}] \int_{\gamma_i} I_{\gamma_i} [1-t^{\gamma_j}] \int_{\gamma_j} I_{\gamma_j}  \\
& = \int_{\gamma_i} I_{\gamma_i}   \int_{\gamma_j} I_{\gamma_j} -t^{\gamma_i}\int_{\gamma_i} I_{\gamma_i} \int_{\gamma_j} I_{\gamma_j}  -\int_{\gamma_i} I_{\gamma_i}  t^{\gamma_j} \int_{\gamma_j} I_{\gamma_j}  +t^{\gamma_i} \int_{\gamma_i} I_{\gamma_i} t^{\gamma_j} \int_{\gamma_j} I_{\gamma_j}  \\
\end{split}
\ee
As before, after discarding, we have the forests
\be
 U_{\text{disj}}=   \{ \gamma_i, \gamma_j\}, \{\gamma_i \}, \{\gamma_j \}, \emptyset.
\ee
We have seen the example (\ref{w8}), in which we need no extra renormalization for the total amplitude besides the one from the forest $ \{ \gamma_i, \gamma_j\}$. The same forest is obtained when $ \gamma_i$ and $\gamma_j$ are overlapping. In this case, we define the integrands not to include the overlapping line as $I'_{ \gamma_i},I'_{ \gamma_j}, A= \int_{ \gamma_i} \int_{ \gamma_j} I'_{ \gamma_i},I'_{ \gamma_j} \Delta_{\text{overlap}}.$ The proof is essentially the same.

The collection of the forests, called the family, has the structure of a power set,
\be
{\cal F}_{\text{nes}}= 2^{\{\gamma, \gamma_i \}} ,\quad  {\cal F}_{\text{disj}}  = 2^{\{\gamma_i,\gamma_j\}},
\ee 
which is obvious from the subtraction structure in the effective coupling and its resolution. The relation holds if the diagrams are subamplitudes of a bigger diagram. 

We know that an amplitude $A^{(n)}_L$ contains either nested or disjoint subdiagrams from the generation.
Therefore, reversing the logic, we can prove the above relation from the forest. Our method of renormalization gives an equivalent result to the BPHZ scheme  \cite{Bogoliubov:1957gp, Hepp:1966eg, Zimmermann:1967, Zimmermann:1969jj}.
The idea is the same: replace every divergent (sub)diagram with a renormalized one \cite{Bogoliubov:1957gp}, which is justified by the addition of counterterms or the removal of the bare parameter. From our approach, the reason for using the forest is clear (See also \cite{Kreimer:1997dp, Kreimer:1998iv}). 

We consider two more nontrivial cases. First, consider two disjoint graphs $\gamma_i,\gamma_j$ contained. Applying the above procedures, we get the forests
\be
 {\cal F}_{\text{all inside}} = 2^{ \{ \gamma,\gamma_i,\gamma_j \} }, \quad \gamma_i \cap \gamma_j = \emptyset, \quad \gamma_i \cup \gamma_j \subset \gamma.
\ee
Since $\gamma_j \subset \gamma$, the Taylor operators do not commute
\be \label{Tcommute}
 t^{\gamma}t^{\gamma_j} \ne t^{\gamma_j}t^{\gamma},
\ee
ant the LHS is in the correct order.
It should be understood that for $\{\gamma,\gamma_j \}$, we should apply the Taylor operator as $t^\gamma t^{\gamma_i}$ from the nesting structure. This is automatically done if we regard every subset in $\cal F$ by the nested ordering whose operator we denote ${\cal N}$.
The next case is when only $\gamma_i$ is nested in $\gamma$ and $\gamma_j$ is disjoint to both, we have
\be
\{ U_{\gamma^j \text{outside}}\} = 2^{ \{ \gamma,\gamma_i,\gamma_j \} }, \quad \gamma \cap \gamma_j = \emptyset, \quad \gamma_i  \subset \gamma.
\ee
Formally, the answer looks the same. This time, the commutation relation holds
\be
 t^{\gamma}t^{\gamma_j}=t^{\gamma_j}t^{\gamma}.
\ee
Nevertheless, if we obey the ordering of the nested structure, we can apply the Taylor expansion without subtlety. 

Finally, a given amplitude $A^{(n)}_L$ may have overlapping loops in general so that we have more than one way of defining subamplitudes. As we have seen in the generating method, taking all of the possible ``routes'' should give the correct coefficient for the renormalized one $\overline A^{(n)}_L.$ Therefore, we include all the routes in the forest.

This reasoning led us to the following theorem.
\begin{theorem}[family of forests]
The forest $U$ is the nesting-ordered power set of disjoint (sub)diagrams $\gamma_i \subseteq \Gamma$ with $\delta(\gamma_i)\ge 0$
\be \label{family}
{\cal F} = \bigcup_{\text{routes}} U, \quad U = {\cal N} \big( 2^{ \bigcup \{\gamma_i \} } \big),
\ee
where ${\cal N}$ is the nesting-ordering operator.
\end{theorem}

Although the solution is equivalent, there is a little difference. As we have seen in Section \ref{sec:2loop}, we can track the origin of the overlapping divergence. From the example of the sunset diagram, we find that we do not care about the overlapping and just need to modify the symmetric factor. This is consistent with Theorem \ref{thm:bz}, and we do not need to worry about overlapping in applying the forest formula.

Also, technically, our method is easier to use. In the forest formula, we need to find maximal subamplitudes. It is easier to find a small loop than to find a maximal subloop required in the forest formula. After contraction, the corresponding weighted Feynman diagram becomes simpler, on which we iterate the procedures. In our approach, the renormalized amplitude is, if generated, already given with the correct coefficient. So, if we find one route, the renormalization ends.

\subsection{Tracking forests}

Here, we present a method to track the forests. We need to understand the nesting structure of the subamplitudes. Since an irreducible loop is the minimal in $r$-contraction, we find it and contract. This gives us a contracted/weighted diagram. Iterating this gives us a tree diagram in the end. If we remember the order of contraction and the original shapes of the contracted diagrams, we can not only recover the original diagram but also the nesting structure; by checking a previously contracted vertex may belong to another loop.  

To do this, the following recording method for nesting structure will be useful.
\begin{lemma}[the rightmost rule] \label{lemrightmost}
Suppose we have a collection of objects. Assume any two objects are either nested or disjoint. 
An ordered set with parentheses (and commas) can describe the nesting structure of objects.
\begin{enumerate}
\item Inside the common parenthesis, the rightmost element contains all the rest.
\item The rest of the elements are disjoint.
\item If all the elements are disjoint, we make the rightmost entry empty.
\end{enumerate}
\end{lemma}
For instance, five objects $\{a,b,c,d,e\}$ with the relations 
\be a  \subset c \subset e, \quad b \subset c, \quad \quad d \subset e,
\ee
otherwise disjoint, are denoted as
\be
 ((a,b,c),d,e).
\ee
This rule is possible because an element cannot belong to more than one element at the same time.
The disjoint objects can be grouped in the common parenthesis, listed on the left. The rightmost element is indicated by the closing parenthesis. A product of two loops with a common vertex has no total object, as we have seen in Sec \ref{sec:2loop}.

\begin{theorem}[tracking forests] \label{thm:ren}
The following procedure gives forests $U$ for an unrenormalized 1PI amplitude $A_L$.
\end{theorem}
\begin{enumerate}
\item 
We look for and contract an irreducible subdiagram $A_l$, where $l$ is the loop order in the sense of (\ref{bigK}). We $v$-contract this subdiagram $A_l$ to a point $v$
\be 
 A_{l} (v),
\ee
where  $v$ is the location of the contracted $A_l$ as the vertex in $\Gamma/A_l$. 
For example, on each row in Fig. \ref{fig:lambda2}, one of the one-loop subdiagrams on the RHS is contracted to a point, yielding the diagram on the LHS. 

\item
For a propagator, we do the same.
If we have a linear chain of identical 1PI diagrams, as in Fig. \ref{fig:M1}, we contract them at once to make one propagator $v$ as in (\ref{LCMrep}),
\be
A_{(l)}^{(2)}(v) = D_{(l)}^{-1}(v), 
\ee
which is also recorded as one entry. In this case, the order $l_1$ is the order of the 1PI component.

\item
Repeat the above procedures. 
At each step, we obtain the weighted Feynman diagram. If we record the resulting coupling in the ``stack'', that is, a sequence forming an ordered set 
\be \label{stack}
 {\cal P} = \left( A^{(n_1)}_{l_1}(v_1),  A^{(n_2)}_{l_2}(v_2),  A^{(n_3)}_{l_3}(v_3), \dots,  A^{(n_N)}_{l_N}(v_N) \right),
\ee
where the subscript means the order. 

\item
During each procedure, a former subgraph may be included in one of the later subgraphs. Using the rightmost rule, Lemma \ref{lemrightmost}, record the nesting structure. This will add more parentheses in the stack (\ref{stack}). 
This defines the nesting-ordering operator ${\cal N}$ in (\ref{family}).
\item
In the end, we obtain a tree diagram of the total loop order 
\be \sum_{i=1}^N l_i.
\ee 
Now we understand the nested, disjoint or overlapping structure. The family of the forest is
\be \label{thisfamily} {\cal N} (2^{\cal P})
\ee 
with the set structure and ordering ${\cal N}$.
\item 
There can be more than one way to perform the contraction. They commute and we arrive at the same weighted Feynman diagram, containing no loop. We called these routes. Starting from the tree diagram in Procedure 5 and generate the complete vertex function $\overline \Gamma_L^{(n)}$ in which the original diagram is contained. All the intermediate routes give the complete family ${\cal F}$ of the forests.
\end{enumerate}

\subsection{An example}

\begin{figure}[t]
\begin{center}
\includegraphics[scale=0.9]{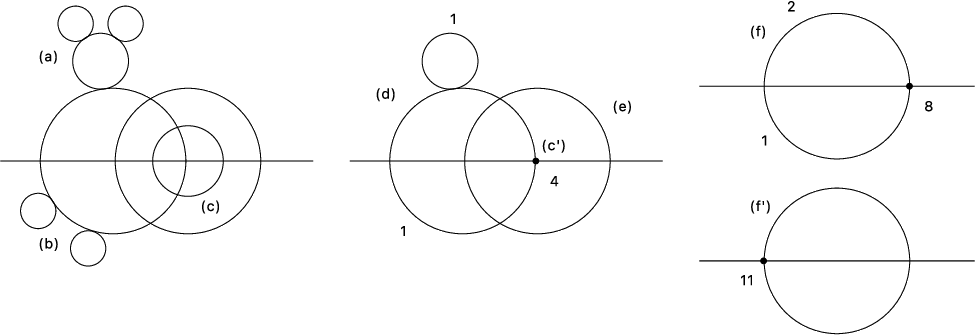}
\end{center}
\caption{Renormalization of a 15-loop diagram. Taking into account 1-particle-reducible (1PR) subdiagrams in the propagators, the effective loop order is 13. After contracting all the subdiagrams, indeed, we obtain an order-13 propagator. Each vertex functions are renormalized. \label{fig:genloop}}
\end{figure}

We take an example in the $\phi^4$-theory. Consider an amplitude with the 15-loop Feynman diagram given in Fig. \ref{fig:genloop}. Taking into account one-particle-reducible subgraphs in the propagators, it is effectively a 13-loop diagram; since it has two scalar external legs, it reduces to a scalar propagator. So, we expect that this appears as one of the graphs in the propagator $D_{(13)}(p^2)$.

For this particular diagram, the renormalization procedure is as follows.
First, we have 1PR parts in (a) and (b), each of which belongs to an order one propagator $D_{(1)}({\rm a})$ and $D_{(1)}({\rm b}),$ respectively. The subdiagram (c) is an irreducible four-loop order with four external legs and belongs to an order four quartic coupling $\lambda_{4}({\rm c})$. 

Contracting all of these, we get the second diagram in Fig. \ref{fig:genloop}.
The part (d) subdiagram, a one-loop-corrected propagator containing $D_{(1)}({\rm a})$, becomes the part (f) subdiagram in the last diagram $D_{(2)}({\rm d})$, after contracting the loop. The subdiagram (e) is again an irreducible diagram of order 4, containing the order four coupling $\lambda_{8}({\rm c}')$. Therefore, with the order four vertex in the 4-loop diagram, we have an order $4+4=8$ vertex $\lambda_{8}({\rm e})$ as in the last diagram.

Contracting all, we obtain the last diagram that has the same structure as the sunset diagram. We have an order two propagator and an order one propagator in part  
order 11 vertex, forming a 2-loop diagram. Contracting fish diagrams twice, we obtain an order-13 propagator $D_{(13)}$.

Alternatively, from the middle diagram, we may take another contraction to obtain a different vertex function.
For example, we may contact the subdiagram (d) first. The upper loop makes an effective propagator of order 2. The resulting quartic coupling has the order $1+2+4+4 = 11$. With this, the final diagram is another sunset (2-loop) diagram with unmodified propagators. It can be contracted to make the order 13-diagram for $D_{(13)}(p^2)$.
The above process records the nesting structure of the diagrams as
\be
 {\cal P} = \Big( \big((D_{(1)}({\rm a}),D_{(2)}({\rm d})), D_{(1)}({\rm b}), (\lambda_{4}({\rm c}),\lambda_{8}({\rm c'})), D_{(11)} \big), D_{(13)} \Big).
\ee
The family of forests we just found is (\ref{thisfamily}).

\section{Discussion}

We presented a new (and old) view of renormalization that makes use of the self-similarity of amplitudes. A loop amplitude $\overline A_L^{(n)}(\p)$ that is fully renormalized has self-similarity in the sense that it also plays the role of a coupling inside another amplitude or even itself. This means the sum of these in various loop order acts as an effective coupling, similar to the coupling in a free field theory.

Such an object is best described by the vertex functions $\overline \Gamma^{(n)}(\p)$, the coefficients of the effective potential, including perturbative corrections. They are ``physical'' because they appear in the $S$-matrix and with them, we can test the theory in the experiments. The success of QFT is its prediction on the scale dependence of $\overline \Gamma^{(n)}(\p)$, in which the scale is specified by the set of the external momenta $\p$. In other words, we obtain how the correction changes the value of effectively the same coupling across scales. Loop correction modifies the coupling whose relevance depends on the scale $\p$. In renormalizable theory, a renormalized quantity has the subtraction (Taylor expansion) structure so that it is finite, independent of regularization and insensitive to cutoff, provided decoupling takes place. Otherwise, the unrenormalized amplitude is divergent and cannot become a small coupling admitting perturbation.

Having a well-defined unit of physical parameters as the fully renormalized effective couplings, the following features of QFT are clearly seen in the effective action. First, all the renormalization we use is finite, relating two couplings across a finite range as in (\ref {finiteren}) and the cancellation of divergence is a byproduct. The issue of infinity only lies in the bare parameter in the Lagrangian. Every quantity appearing in the effective action can be written in terms of observable instead of the bare ones, as Theorem \ref{thm:noinf} shows. 
Also, we use relative physical parameter $\overline \Gamma^{(n)}_{(L)}(\p)$ from the reference parameter defined at a scale. The scale is specified by the external momenta $\p$ and it should be kept in the effective action to renormalize the couplings. This is the prediction of QFT.

With this, we can systematically generate higher-loop corrected amplitudes that are fully renormalized. 
The generation process includes the complete renormalization. It is not merely extracting the finite part but obtaining a physical quantity. To do this, we need the sum $\overline \Gamma_{(L)}^{(n)}(\p)$ of complete couplings belonging to the same $n$-point correlation function up to the order $L$ that we want to calculate. It is because the renormalization (\ref{effcoup}) condition to be fixed by the observed quantity at the reference scale, $\overline \Gamma_0^{(n)}(\p),$ needs the full summation of couplings to the order we can access. 

The generation method helps us understand overlapping divergences. An amplitude for overlapping loops can be made asymmetrical by contracting to a loop amplitude without overlapping divergence. Even if we treat them as nested diagrams, the symmetric combination of the possible diagrams is renormalized as if the two are disjoint. This explains the result of Berg\`ere and Zuber, Theorem \ref{thm:bz}, and the reason we should make a forest without overlapping diagrams in the BPHZ renormalization. A possible improvement would be to find a good way to automatically implement the symmetric factor if the diagram after resolution acquires enhanced symmetry.

In the end, the only quantities we can observe via the $S$-matrix are these fully renormalized couplings $\overline \Gamma^{(n)}(\p)$. 
For instance, neither $m^2_B,\tilde \Sigma(p^2)$ or even the mass (\ref{SlidingMass}) is separately observable \cite{Choi:2023cqs}. In fact, the order two coupling $\overline \Gamma^{(2)}(p^2) = p^2+m^2(p^2)$ or its inverse is the minimal unit appearing in the $S$-matrix.
As an important application, the Higgs mass correction as a function of the external momentum can be compared with the experimental data \cite{Choi:2023cqs}. The act of renormalization not only removes the divergence but also imposes the momentum dependence on the coupling \cite{Choi:2024cbs}. 

The uniqueness of the effective coupling in a given theory tells us what physical quantity we should compare with the experiment. We can save, retrieve, and share the results, and we can do numerical calculations fully or partly. 

We can also prove the renormalizability of a given diagram, which is equivalent but an alternative description to BPHZ. 
Although our framework provides a good alternative way to solve the Bogoliubov recursion relation for renormalization, rather than performing the renormalization procedure, it is more appropriate to generate the diagram, giving the process that we are interested in. In other words, any given $L$-loop diagram is obtained from the order-$L$ coupling with the same external legs. In practice, we are not interested in calculating a particular amplitude, but we want the sum of all diagrams of a given order.

We distinguish the three concepts sharing the same name of renormalization. The field-strength re-normalization (FSR) is the origin of the term. However, the above Kadanoffian regards the loop-corrected quantity as an effective quantity as in the free field. We call this self-similar renormalization (SSR). Finally, we will show that the process of removing divergence is, in fact, writing the same quantity in reference to a unit parameter. This we call relative-value rewriting (RVR).

It would be interesting to have a general method to generate all the irreducible diagrams exhaustively. The elements of resolution and generations are irreducible amplitudes. They are present at sufficiently high orders, but their patterns seem irregular. For this, we have provided a search method using flop change, however, it is not efficient. Understanding algebra behind Feynman diagrams would help \cite{Kreimer:1997dp, Kreimer:1998iv}.

\subsection*{Acknowledgments}

This work is partly supported by the grant RS-2023-00277184 of the National Research Foundation of Korea. The author is grateful to Hyun-Min Lee for the discussions, to Michael Ratz, Yuri Shirman and Timothy Tait for the comments and to the Particle Physics Group of UC Irvine and the Theory Group of Chung-Ang University and hospitality while the work was finalized.


\begin{thebibliography}{99}

\bibitem{Kadanoff:1966wm}
L.~P.~Kadanoff,
``Scaling laws for Ising models near T(c),''
Physics Physique Fizika \textbf{2} (1966), 263-272
doi:10.1103/PhysicsPhysiqueFizika.2.263


\bibitem{Nambu:1968rr}
Y.~Nambu,
``S Matrix in semiclassical approximation,''
Phys. Lett. B \textbf{26} (1968), 626-629
doi:10.1016/0370-2693(68)90436-X;

S.~R.~Coleman and E.~J.~Weinberg,
``Radiative Corrections as the Origin of Spontaneous Symmetry Breaking,''
Phys. Rev. D \textbf{7} (1973), 1888-1910
doi:10.1103/PhysRevD.7.1888



\bibitem{Choi:2023cqs}
K.~S.~Choi,
J. Korean Phys. Soc. \textbf{84} (2024) no.8, 591-595
[erratum: J. Korean Phys. Soc. \textbf{86} (2025) no.2, 156]
doi:10.1007/s40042-024-01025-7
[arXiv:2310.00586 [hep-ph]].

\bibitem{Choi:2023mma}
K.~S.~Choi,
Phys. Rev. D \textbf{109} (2024) no.7, 076008
doi:10.1103/PhysRevD.109.076008
[arXiv:2310.10004 [hep-th]].

\bibitem{Choi:2024hkd}
K.~S.~Choi,
``Stability of the Scalar Mass against Loop Corrections,''
[arXiv:2410.21118 [hep-ph]].


\bibitem{Choi:2024cbs}
K.-S.~Choi,
``Renormalization, Decoupling and the Hierarchy Problem,''
[arXiv:2408.06406 [hep-ph]].

\bibitem{Kleinert:1982ki}
H.~Kleinert,
``HIGHER EFFECTIVE ACTIONS FOR BOSE SYSTEMS,''
Fortsch. Phys. \textbf{30} (1982), 187-232
doi:10.1002/prop.19820300402

\bibitem{Kastening:1999fy}
B.~M.~Kastening,
``Recursive graphical construction of Feynman diagrams in phi**4 theory: Asymmetric case and effective energy,''
Phys. Rev. E \textbf{61} (2000), 3501-3528
doi:10.1103/PhysRevE.61.3501
[arXiv:hep-th/9908172 [hep-th]].

\bibitem{Kleinert:1999uv}
H.~Kleinert, A.~Pelster, B.~M.~Kastening and M.~Bachmann,
``Recursive graphical construction of Feynman diagrams and their multiplicities in phi**4 theory and in phi**2 A theory,''
Phys. Rev. E \textbf{62} (2000), 1537-1559
doi:10.1103/PhysRevE.62.1537
[arXiv:hep-th/9907168 [hep-th]].

\bibitem{Ilderton:2005vg}
A.~Ilderton,
``QED amplitudes: Recurrence relations to all orders,''
Nucl. Phys. B \textbf{742} (2006), 176-186
doi:10.1016/j.nuclphysb.2006.02.040
[arXiv:hep-th/0512007 [hep-th]].

\bibitem{Choi:2007nb}
K.~S.~Choi and T.~Kobayashi,
``Higher order couplings from heterotic orbifold theory,''
Nucl. Phys. B \textbf{797} (2008), 295-321
doi:10.1016/j.nuclphysb.2008.01.016
[arXiv:0711.4894 [hep-th]].


\bibitem{Kim:2008hda}
C.~Kim,
``Recursion Relation for the Feynman Diagrams of the Effective Action for the Third Legendre Transformation,''
J. Korean Phys. Soc. \textbf{53} (2008), 3170
doi:10.3938/jkps.53.3164
[arXiv:0807.2305 [hep-th]].

\bibitem{Abe:2009dr}
H.~Abe, K.~S.~Choi, T.~Kobayashi and H.~Ohki,
``Higher Order Couplings in Magnetized Brane Models,''
JHEP \textbf{06} (2009), 080
doi:10.1088/1126-6708/2009/06/080
[arXiv:0903.3800 [hep-th]].




\bibitem{Borinsky:2022lds}
M.~Borinsky and O.~Schnetz,
``Recursive computation of Feynman periods,''
JHEP \textbf{22} (2022), 291
doi:10.1007/JHEP08(2022)291
[arXiv:2206.10460 [hep-th]].

\bibitem{Polchinski:1983gv}
J.~Polchinski,
``Renormalization and Effective Lagrangians,''
Nucl. Phys. B \textbf{231} (1984), 269-295
doi:10.1016/0550-3213(84)90287-6

\bibitem{Bogoliubov:1957gp}
N.~Bogoliubov and O.~Parasiuk, 
``Uber die Multiplikation der Kausalfunktionen in der Quantentheorie der Felder,''
Acta Math. \textbf{97}
  (1957) 227--266.

\bibitem{Hepp:1966eg}
K.~Hepp,
``Proof of the Bogolyubov-Parasiuk theorem on   renormalization,'' 
Commun. Math. Phys. \textbf{2} (1966) 301--326.

\bibitem{Zimmermann:1967}
W. Zimmermann,
``Local field equation for $A^{4}$-coupling in renormalized perturbation theory,''
Commun. Math. Phys. \textbf{6} (1967) 161--188.

\bibitem{Zimmermann:1969jj}
W.~Zimmermann,
``Convergence of Bogolyubov's method of renormalization in momentum space,''
Commun. Math. Phys. \textbf{15}, 208-234 (1969)
doi:10.1007/BF01645676;

W.~Zimmermann, in 
S.~Deser, M.~Grisaru, H.~Pendleton, Lectures on Elementary Particles and Quantum Field Theory. Volume 1. 1970 Brandeis University Summer Institute in Theoretical Physics, 1970.

\bibitem{Lowenstein:1975rg}
J.~H.~Lowenstein and W.~Zimmermann,
``The Power Counting Theorem for Feynman Integrals with Massless Propagators,''
Commun. Math. Phys. \textbf{44} (1975), 73-86
doi:10.1007/BF01609059

\bibitem{Lowenstein:1975ps}
J.~H.~Lowenstein,
``Convergence Theorems for Renormalized Feynman Integrals with Zero-Mass Propagators,''
Commun. Math. Phys. \textbf{47} (1976), 53-68
doi:10.1007/BF01609353

\bibitem{BS}
Bogoliubov, D. V. Shirkov,  ``Introduction to the Theory of Quantized Fields.''   John Wiley \& Sons Inc; 3rd edition (1980).

\bibitem{Sibold}
Klaus Sibold, ``Bogoliubov-Parasiuk-Hepp-Zimmermann renormalization scheme,'' Scholarpedia, 5(5), (2010), 7306.

\bibitem{Blaschke:2013cba}
D.~N.~Blaschke, F.~Gieres, F.~Heindl, M.~Schweda and M.~Wohlgenannt,
``BPHZ renormalization and its application to non-commutative field theory,''
Eur. Phys. J. C \textbf{73} (2013), 2566
doi:10.1140/epjc/s10052-013-2566-8
[arXiv:1307.4650 [hep-th]].

\bibitem{Herzog:2017jgk}
F.~Herzog,
Nucl. Phys. B \textbf{926} (2018), 370-380
doi:10.1016/j.nuclphysb.2017.11.011
[arXiv:1711.06121 [hep-ph]].



\bibitem{Collins:1984xc}
J.~C.~Collins,
``Renormalization,'
Cambridge University Press, 2023,
ISBN 978-0-521-31177-9, 978-0-511-86739-2, 978-1-00-940180-7, 978-1-00-940176-0, 978-1-00-940179-1
doi:10.1017/9781009401807

\bibitem{Itzykson:1980rh}
C.~Itzykson and J.~B.~Zuber,
McGraw-Hill, 1980,
ISBN 978-0-486-44568-7


\bibitem{Duncan:2012aja}
A.~Duncan,
``The Conceptual Framework of Quantum Field Theory,''
Oxford University Press, 2012,
ISBN 978-0-19-880765-0, 978-0-19-880765-0, 978-0-19-957326-4
doi:10.1093/acprof:oso/9780199573264.001.0001


\bibitem{Buchbinder:2021wzv}
I.~L.~Buchbinder and I.~Shapiro,
``Introduction to Quantum Field Theory with Applications to Quantum Gravity,''
Oxford University Press, 2023,
ISBN 978-0-19-887234-4, 978-0-19-883831-9
doi:10.1093/oso/9780198838319.001.0001

\bibitem{Talagrand:2022huy}
M.~Talagrand,
``What Is a Quantum Field Theory?,''
Cambridge University Press, 2022,
ISBN 978-1-108-22514-4, 978-1-316-51027-8
doi:10.1017/9781108225144

\bibitem{Sirlin:1980nh}
A.~Sirlin,
``Radiative Corrections in the SU(2)-L x U(1) Theory: A Simple Renormalization Framework,''
Phys. Rev. D \textbf{22} (1980), 971-981
doi:10.1103/PhysRevD.22.971

A.~Denner,
``Techniques for calculation of electroweak radiative corrections at the one loop level and results for W physics at LEP-200,''
Fortsch. Phys. \textbf{41} (1993), 307-420
doi:10.1002/prop.2190410402
[arXiv:0709.1075 [hep-ph]].

\bibitem{Weinberg:1959nj}
S.~Weinberg,
``High-energy behavior in quantum field theory,''
Phys. Rev. \textbf{118}, 838-849 (1960)
doi:10.1103/PhysRev.118.838;

\bibitem{Peskin:1995ev}
M.~E.~Peskin and D.~V.~Schroeder,
``An Introduction to quantum field theory,''
Addison-Wesley, 1995,
ISBN 978-0-201-50397-5


\bibitem{Coleman:2018mew}
S.~Coleman, B.~G.~g.~Chen, D.~Derbes, D.~Griffiths, B.~Hill, R.~Sohn and Y.~S.~Ting,
``Lectures of Sidney Coleman on Quantum Field Theory,''
WSP, 2018,
ISBN 978-981-4632-53-9, 978-981-4635-50-9
doi:10.1142/9371


\bibitem{Kleinert:2001ax}
H.~Kleinert and V.~Schulte-Frohlinde,
``Critical properties of phi**4-theories,''
(2001), World Scinetific

\bibitem{Appelquist:1974tg}
T.~Appelquist and J.~Carazzone,
``Infrared Singularities and Massive Fields,''
Phys. Rev. D \textbf{11}, 2856 (1975)
doi:10.1103/PhysRevD.11.2856;

K.~Symanzik,
``Infrared singularities and small distance behavior analysis,''
Commun. Math. Phys. \textbf{34} (1973), 7-36
doi:10.1007/BF01646540

\bibitem{Jackiw:1974cv}
R.~Jackiw,
``Functional evaluation of the effective potential,''
Phys. Rev. D \textbf{9} (1974), 1686
doi:10.1103/PhysRevD.9.1686

\bibitem{Weinberg:1995mt}
S.~Weinberg,
``The Quantum theory of fields. Vol. 1: Foundations,''
Cambridge University Press, 2005,
ISBN 978-0-521-67053-1, 978-0-511-25204-4
doi:10.1017/CBO9781139644167

\bibitem{Bielas:2013rja}
K.~Bielas, I.~Dubovyk, J.~Gluza and T.~Riemann,
``Some Remarks on Non-planar Feynman Diagrams,''
Acta Phys. Polon. B \textbf{44} (2013) no.11, 2249-2255
doi:10.5506/APhysPolB.44.2249
[arXiv:1312.5603 [hep-ph]].

\bibitem{Dyson:1949ha}
F.~J.~Dyson,
``The S matrix in quantum electrodynamics,''
Phys. Rev. \textbf{75} (1949), 1736-1755
doi:10.1103/PhysRev.75.1736

A.~Salam,
``Overlapping divergences and the S matrix,''
Phys. Rev. \textbf{82} (1951), 217-227
doi:10.1103/PhysRev.82.217

A.~Salam,
``Divergent integrals in renormalizable field theories,''
Phys. Rev. \textbf{84} (1951), 426-431
doi:10.1103/PhysRev.84.426

\bibitem{Bergere:1974zh}
M.~C.~Bergere and J.~B.~Zuber,
``Renormalization of Feynman amplitudes and parametric integral representation,''
Commun. Math. Phys. \textbf{35} (1974), 113-140
doi:10.1007/BF01646611

\bibitem{Kreimer:1997dp}
D.~Kreimer,
``On the Hopf algebra structure of perturbative quantum field theories,''
Adv. Theor. Math. Phys. \textbf{2} (1998), 303-334
doi:10.4310/ATMP.1998.v2.n2.a4
[arXiv:q-alg/9707029 [math.QA]].


\bibitem{Kreimer:1998iv}
D.~Kreimer,
``On overlapping divergences,''
Commun. Math. Phys. \textbf{204} (1999), 669
doi:10.1007/s002200050661
[arXiv:hep-th/9810022 [hep-th]].

\end{thebibliography}
\end{document}